\documentclass[
aps,
prd,
twocolumn,
superscriptaddress,
showpacs
]{revtex4-1}
\usepackage{graphicx}
\usepackage{amssymb}
\usepackage{color}
\usepackage{xspace}
\usepackage{cleveref}
\usepackage{multirow}

\newcommand{\nue}{\nu_e}
\newcommand{\numu}{\nu_\mu}

\newcommand{\nuebar}{\bar{\nu}_e}
\newcommand{\numubar}{\bar{\nu}_\mu}

\newcommand{\degr}{\ensuremath{^{\circ}}\xspace}

\newcommand{\val}[2]{\ensuremath{#1 \; \mathrm{#2}\xspace}} 

\newcommand{\atom}[2]{\ensuremath{{}^{#1}\textrm{#2}}\xspace}
\newcommand{\atome}[2]{\ensuremath{{}^{#1}\textrm{#2}^*}\xspace}

\newcommand{\done}{\ensuremath{\mathcal{D}_{1}}\xspace}
\newcommand{\dtwo}{\ensuremath{\mathcal{D}_{2}}\xspace}
\newcommand{\gt}{\ensuremath{g_t}\xspace}
\newcommand{\gp}{\ensuremath{g_p}\xspace}
\newcommand{\q}{\ensuremath{Q_{\rm{rec}}}\xspace}

\newcommand{\ncoth}{NC non-QE\xspace}

\newcommand{\gmr}{$\gamma$ ray\xspace}
\newcommand{\gmrs}{$\gamma$ rays\xspace}
\newcommand{\cc}{\mathit{c}}

\newcommand{\fluxmean}{630 MeV\xspace}
\newcommand{\fluxtype}{median\xspace}

\newcommand{\ncel}{34.8\xspace}
\newcommand{\bkgd}{16.2\xspace}

\newcommand{\bkgdnon}{1.2\xspace}
\newcommand{\obs}{43\xspace}
\newcommand{\mctot}{51.0\xspace}
\newcommand{\xsec}{1.55\xspace}
\newcommand{\xsecpre}{2.01\xspace}
\newcommand{\xseclow}{1.20\xspace}
\newcommand{\xsechigh}{2.26\xspace}
\newcommand{\xseclowninety}{0.96\xspace}
\newcommand{\xsechighninety}{2.78\xspace}

\graphicspath{{fig/}}

\begin{document}


\title{Measurement of the neutrino-oxygen neutral-current interaction cross section by observing nuclear deexcitation $\gamma$ rays}

\newcommand{\INSTC}{\affiliation{University of Alberta, Centre for Particle Physics, Department of Physics, Edmonton, Alberta, Canada}}
\newcommand{\INSTEE}{\affiliation{University of Bern, Albert Einstein Center for Fundamental Physics, Laboratory for High Energy Physics (LHEP), Bern, Switzerland}}
\newcommand{\INSTFE}{\affiliation{Boston University, Department of Physics, Boston, Massachusetts, U.S.A.}}
\newcommand{\INSTD}{\affiliation{University of British Columbia, Department of Physics and Astronomy, Vancouver, British Columbia, Canada}}
\newcommand{\INSTGA}{\affiliation{University of California, Irvine, Department of Physics and Astronomy, Irvine, California, U.S.A.}}
\newcommand{\INSTI}{\affiliation{IRFU, CEA Saclay, Gif-sur-Yvette, France}}
\newcommand{\INSTGB}{\affiliation{University of Colorado at Boulder, Department of Physics, Boulder, Colorado, U.S.A.}}
\newcommand{\INSTFG}{\affiliation{Colorado State University, Department of Physics, Fort Collins, Colorado, U.S.A.}}
\newcommand{\INSTFH}{\affiliation{Duke University, Department of Physics, Durham, North Carolina, U.S.A.}}
\newcommand{\INSTBA}{\affiliation{Ecole Polytechnique, IN2P3-CNRS, Laboratoire Leprince-Ringuet, Palaiseau, France }}
\newcommand{\INSTEF}{\affiliation{ETH Zurich, Institute for Particle Physics, Zurich, Switzerland}}
\newcommand{\INSTEG}{\affiliation{University of Geneva, Section de Physique, DPNC, Geneva, Switzerland}}
\newcommand{\INSTDG}{\affiliation{H. Niewodniczanski Institute of Nuclear Physics PAN, Cracow, Poland}}
\newcommand{\INSTCB}{\affiliation{High Energy Accelerator Research Organization (KEK), Tsukuba, Ibaraki, Japan}}
\newcommand{\INSTED}{\affiliation{Institut de Fisica d'Altes Energies (IFAE), Bellaterra (Barcelona), Spain}}
\newcommand{\INSTEC}{\affiliation{IFIC (CSIC \& University of Valencia), Valencia, Spain}}
\newcommand{\INSTEI}{\affiliation{Imperial College London, Department of Physics, London, United Kingdom}}
\newcommand{\INSTGF}{\affiliation{INFN Sezione di Bari and Universit\`a e Politecnico di Bari, Dipartimento Interuniversitario di Fisica, Bari, Italy}}
\newcommand{\INSTBE}{\affiliation{INFN Sezione di Napoli and Universit\`a di Napoli, Dipartimento di Fisica, Napoli, Italy}}
\newcommand{\INSTBF}{\affiliation{INFN Sezione di Padova and Universit\`a di Padova, Dipartimento di Fisica, Padova, Italy}}
\newcommand{\INSTBD}{\affiliation{INFN Sezione di Roma and Universit\`a di Roma ``La Sapienza'', Roma, Italy}}
\newcommand{\INSTEB}{\affiliation{Institute for Nuclear Research of the Russian Academy of Sciences, Moscow, Russia}}
\newcommand{\INSTHA}{\affiliation{Kavli Institute for the Physics and Mathematics of the Universe (WPI), Todai Institutes for Advanced Study, University of Tokyo, Kashiwa, Chiba, Japan}}
\newcommand{\INSTCC}{\affiliation{Kobe University, Kobe, Japan}}
\newcommand{\INSTCD}{\affiliation{Kyoto University, Department of Physics, Kyoto, Japan}}
\newcommand{\INSTEJ}{\affiliation{Lancaster University, Physics Department, Lancaster, United Kingdom}}
\newcommand{\INSTFC}{\affiliation{University of Liverpool, Department of Physics, Liverpool, United Kingdom}}
\newcommand{\INSTFI}{\affiliation{Louisiana State University, Department of Physics and Astronomy, Baton Rouge, Louisiana, U.S.A.}}
\newcommand{\INSTJ}{\affiliation{Universit\'e de Lyon, Universit\'e Claude Bernard Lyon 1, IPN Lyon (IN2P3), Villeurbanne, France}}
\newcommand{\INSTCE}{\affiliation{Miyagi University of Education, Department of Physics, Sendai, Japan}}
\newcommand{\INSTDF}{\affiliation{National Centre for Nuclear Research, Warsaw, Poland}}
\newcommand{\INSTFJ}{\affiliation{State University of New York at Stony Brook, Department of Physics and Astronomy, Stony Brook, New York, U.S.A.}}
\newcommand{\INSTGJ}{\affiliation{Okayama University, Department of Physics, Okayama, Japan}}
\newcommand{\INSTCF}{\affiliation{Osaka City University, Department of Physics, Osaka, Japan}}
\newcommand{\INSTGG}{\affiliation{Oxford University, Department of Physics, Oxford, United Kingdom}}
\newcommand{\INSTBB}{\affiliation{UPMC, Universit\'e Paris Diderot, CNRS/IN2P3, Laboratoire de Physique Nucl\'eaire et de Hautes Energies (LPNHE), Paris, France}}
\newcommand{\INSTGC}{\affiliation{University of Pittsburgh, Department of Physics and Astronomy, Pittsburgh, Pennsylvania, U.S.A.}}
\newcommand{\INSTFA}{\affiliation{Queen Mary University of London, School of Physics and Astronomy, London, United Kingdom}}
\newcommand{\INSTE}{\affiliation{University of Regina, Department of Physics, Regina, Saskatchewan, Canada}}
\newcommand{\INSTGD}{\affiliation{University of Rochester, Department of Physics and Astronomy, Rochester, New York, U.S.A.}}
\newcommand{\INSTBC}{\affiliation{RWTH Aachen University, III. Physikalisches Institut, Aachen, Germany}}
\newcommand{\INSTFB}{\affiliation{University of Sheffield, Department of Physics and Astronomy, Sheffield, United Kingdom}}
\newcommand{\INSTDI}{\affiliation{University of Silesia, Institute of Physics, Katowice, Poland}}
\newcommand{\INSTEH}{\affiliation{STFC, Rutherford Appleton Laboratory, Harwell Oxford,  and  Daresbury Laboratory, Warrington, United Kingdom}}
\newcommand{\INSTCH}{\affiliation{University of Tokyo, Department of Physics, Tokyo, Japan}}
\newcommand{\INSTBJ}{\affiliation{University of Tokyo, Institute for Cosmic Ray Research, Kamioka Observatory, Kamioka, Japan}}
\newcommand{\INSTCG}{\affiliation{University of Tokyo, Institute for Cosmic Ray Research, Research Center for Cosmic Neutrinos, Kashiwa, Japan}}
\newcommand{\INSTGI}{\affiliation{Tokyo Metropolitan University, Department of Physics, Tokyo, Japan}}
\newcommand{\INSTF}{\affiliation{University of Toronto, Department of Physics, Toronto, Ontario, Canada}}
\newcommand{\INSTB}{\affiliation{TRIUMF, Vancouver, British Columbia, Canada}}
\newcommand{\INSTG}{\affiliation{University of Victoria, Department of Physics and Astronomy, Victoria, British Columbia, Canada}}
\newcommand{\INSTDJ}{\affiliation{University of Warsaw, Faculty of Physics, Warsaw, Poland}}
\newcommand{\INSTDH}{\affiliation{Warsaw University of Technology, Institute of Radioelectronics, Warsaw, Poland}}
\newcommand{\INSTFD}{\affiliation{University of Warwick, Department of Physics, Coventry, United Kingdom}}
\newcommand{\INSTGE}{\affiliation{University of Washington, Department of Physics, Seattle, Washington, U.S.A.}}
\newcommand{\INSTGH}{\affiliation{University of Winnipeg, Department of Physics, Winnipeg, Manitoba, Canada}}
\newcommand{\INSTEA}{\affiliation{Wroclaw University, Faculty of Physics and Astronomy, Wroclaw, Poland}}
\newcommand{\INSTH}{\affiliation{York University, Department of Physics and Astronomy, Toronto, Ontario, Canada}}

\INSTC
\INSTEE
\INSTFE
\INSTD
\INSTGA
\INSTI
\INSTGB
\INSTFG
\INSTFH
\INSTBA
\INSTEF
\INSTEG
\INSTDG
\INSTCB
\INSTED
\INSTEC
\INSTEI
\INSTGF
\INSTBE
\INSTBF
\INSTBD
\INSTEB
\INSTHA
\INSTCC
\INSTCD
\INSTEJ
\INSTFC
\INSTFI
\INSTJ
\INSTCE
\INSTDF
\INSTFJ
\INSTGJ
\INSTCF
\INSTGG
\INSTBB
\INSTGC
\INSTFA
\INSTE
\INSTGD
\INSTBC
\INSTFB
\INSTDI
\INSTEH
\INSTCH
\INSTBJ
\INSTCG
\INSTGI
\INSTF
\INSTB
\INSTG
\INSTDJ
\INSTDH
\INSTFD
\INSTGE
\INSTGH
\INSTEA
\INSTH

\author{K.\,Abe}\INSTBJ
\author{J.\,Adam}\INSTFJ
\author{H.\,Aihara}\INSTCH\INSTHA
\author{T.\,Akiri}\INSTFH
\author{C.\,Andreopoulos}\INSTEH
\author{S.\,Aoki}\INSTCC
\author{A.\,Ariga}\INSTEE
\author{T.\,Ariga}\INSTEE
\author{S.\,Assylbekov}\INSTFG
\author{D.\,Autiero}\INSTJ
% ...

\collaboration{The T2K Collaboration}\noaffiliation

\begin{abstract}
We report the first measurement of the neutrino-oxygen neutral-current quasielastic (NCQE) cross section. It is obtained by observing nuclear deexcitation $\gamma$-rays which follow neutrino-oxygen interactions at the Super-Kamiokande water Cherenkov detector. We use T2K data corresponding to $3.01 \times 10^{20}$ protons on target.  By selecting only events during the T2K beam window and with well-reconstructed vertices in the fiducial volume, the large background rate from natural radioactivity is dramatically reduced.  We observe 43 events in the $4-30$ MeV reconstructed energy window, compared with an expectation of 51.0, which includes an estimated 16.2 background events.  The background is primarily nonquasielastic neutral-current interactions and has only 1.2 events from natural radioactivity. The flux-averaged NCQE cross section we measure is $1.55 \times 10^{-38}$ cm$^2$ with a 68\% confidence interval of $(1.22, 2.20) \times 10^{-38}$ cm$^2$ at a median neutrino energy of 630 MeV, compared with the theoretical prediction of $2.01 \times 10^{-38}$ cm$^2$.
\end{abstract}

\pacs{25.30.Pt, 29.40.Ka, 21.10.Pc, 23.20.Lv}


\maketitle

\section{Introduction}

Nuclear deexcitation \gmrs are a useful tool for detecting neutrino-nucleus neutral-current (NC) interactions where the final state neutrino and associated nucleon are not observed in a Cherenkov detector.  These interactions have previously been observed in neutrino-carbon interactions~\cite{Bodmann:1992ur,Armbruster:1998gk}.  The most well known \gmr production process on oxygen is inelastic scattering, $\nu + \atom{16}{O}\rightarrow \nu + \atome{16}{O}$, where the residual oxygen nucleus can de-excite by emitting a nucleon or \gmrs with energies between \val{1-10}{MeV}. This process can be used to detect supernova neutrinos~\cite{Langanke:1995he}, which have an average energy of \val{20-30}{MeV}. Most theoretical work on \gmr production in NC interactions has been performed in this low neutrino energy range with the assumption that it is applicable up to neutrino energies of several hundred MeV~\cite{Kolbe:1992xu,Beacom:1998ya,*Beacom:1998yb,Nussinov:2000qc,Kolbe:2002gk}.

A recent calculation of \gmr production in neutrino NC interactions shows that quasielastic (QE) nucleon knockout, $\nu + \atom{16}{O} \rightarrow \nu + p + \atome{15}{N}\ (\nu + n + \atome{15}{O})$ overwhelms the inelastic process at $E_{\nu} \gtrsim \val{200}{MeV}$~\cite{Ankowski:2011ei}. The NCQE cross section is calculated to be more than an order of magnitude larger than the NC inelastic cross section from~\cite{Kolbe:2002gk} at $E_{\nu} \approx \val{500}{MeV}$.  
We can observe this interaction in two ways: by observing the `primary' \gmrs produced when the residual nucleus de-excites or by observing the `secondary' \gmrs produced when knocked-out nucleons interact with other nuclei in the water.  The primary \gmrs only occur when the knocked out nucleon comes from the $1p_{3/2}$ or $1s_{1/2}$ states while the secondary \gmrs can be produced by nucleons released during deexcitation or by interactions of the original knocked out nucleons, and so can occur for all nucleon states.
Both types of \gmrs, produced in interactions of atmospheric neutrinos, are a major background for the study of astrophysical neutrinos in the \val{10}{MeV} range~\cite{Bays:2011si,Ueno:2012md} and a direct measurement of the rate of this process with a known neutrino source will be useful for ongoing and proposed projects~\cite{Fukuda:2002uc,Beacom:2003nk,Abe:2011ts,Akiri:2011dv}.

This paper reports the first measurement of the neutrino-oxygen NCQE cross section via the detection of deexcitation \gmrs from both primary and secondary interactions. The neutrinos are produced using the narrow-band neutrino beam at J-PARC and measured with the Super-Kamiokande (SK) water Cherenkov detector.

\section{The T2K Experiment}

The Tokai-to-Kamioka (T2K) experiment~\cite{Abe:2011ks} is a long-baseline neutrino oscillation experiment consisting of a neutrino beam, several near detectors, and using Super Kamiokande as a far detector. It is designed to search for $\numu \to \nue$ appearance, which is sensitive to the neutrino mixing angle $\theta_{13}$, and to precisely measure the mixing angle $\theta_{23}$ and the mass difference $|\Delta m_{32}^{2}|$ by $\numu$ disappearance. 

The accelerator at the Japan Proton Accelerator Research Complex (J-PARC) provides a 30 GeV proton beam which collides with a graphite target to produce charged mesons. Positively-charged pions and kaons are collected and focused by magnetic horns and ultimately decay in flight to produce primarily muon neutrinos inside a \val{96}{m} long cavity filled with helium gas. The proton beam is directed $2.5\degr$ away from SK.  The off-axis neutrino beam has a narrow peak with \fluxtype energy \fluxmean at SK because of the two-body decay kinematics of the $\pi^+$ which dominate the focused beam.  This peak energy was chosen because it corresponds to the first maximum in the neutrino oscillation probability at the location of the far detector. The narrow energy peak also allows for the measurement of the NC cross section at a particular energy. Typically, it is not possible to make energy-dependent measurements of this cross section because the invisible outgoing neutrino makes accurate energy reconstruction impossible.

The T2K experiment has several near detectors located \val{280}{m} from the neutrino production target. The on-axis near detector, INGRID, which consists of 16 modules made up of alternating layers of iron and plastic scintillator arranged in a cross, monitors the neutrino beam direction. 
The off-axis near detectors, ND280, measure the neutrino beam spectrum and composition for the oscillation analyses. The neutrino measurements at the INGRID and ND280 detectors are consistent with expectations~\cite{Abe:2013jth}, but this information is not used to constrain systematic uncertainties in this analysis so that an absolute cross-section measurement can be made.

Super-Kamiokande~\cite{Fukuda:2002uc} is a cylindrical water Cherenkov detector consisting of \val{50}{ktons} of ultra-pure water, located \val{295}{km} from the neutrino target at J-PARC. It was built in the middle of Mt. Ikenoyama, near the town of Kamioka, \val{1000}{m} below the peak. The tank is optically separated into two regions which share the same water. The inner detector (ID) is a cylinder containing the \val{22.5}{kton} fiducial volume and is instrumented with 11,129 inward-facing photomultiplier tubes (PMTs). The outer detector (OD) extends \val{2}{m} outward from all sides of the ID and is instrumented with 1,885 outward-facing PMTs. It  serves as a veto counter against cosmic-ray muons as well as a shield for \gmrs and neutrons emitted from radioactive nuclei in the surrounding rock and stainless steel support structure.

\section{Event simulation}

T2K events at SK are simulated in three stages. First, the neutrino beamline is simulated to predict the flux and energy spectrum of neutrinos arriving at SK. Next, the interactions of those neutrinos with the nuclei in the SK detector are simulated, including final-state interactions within the nucleus. Finally, the SK detector response to all of the particles leaving the nucleus is simulated.

FLUKA~\cite{Battistoni:2011zz} is used to simulate hadron production in the target based on the measured proton beam profile. Hadron production data from NA61/SHINE at CERN~\cite{Abgrall:2011ae,Abgrall:2011ts} is used to tune the simulation and evaluate the systematic error.  The regions of phase space not covered by the NA61/SHINE data directly are tuned by extrapolating the data using the BMPT empirical parameterization~\cite{Bonesini:2001iz}.  Once particles leave the production target they are transported through the magnetic horns, target hall, decay volume, and beam dump using a GEANT3~\cite{Brun:1994aa} simulation with GCALOR~\cite{Zeitnitz:1994bs} for hadronic interactions. The initial composition of the beam is 93\% $\numu$, 6\% $\numubar$, and 1\% $\nue$ (the 0.1\% $\nuebar$ component is not considered in this analysis). The $\numu$ ($\numubar$) flux is 94\% (92\%) from the decay-in-flight of pions, while approximately half of the $\nue$ flux is from muon decays and the other half is from the kaon decays.  A more detailed description of the neutrino flux prediction and its uncertainty can be found in Ref.~\cite{Abe:2012av}. 

\newcommand{\tsp}{\hspace{1em}}
\begin{table}
    \centering
    \begin{tabular}{l@{\tsp}c@{\tsp}c@{\tsp}c}
        \hline\hline 
                                       & $1p_{1/2}$ & $1p_{3/2}$    & $1s_{1/2}$ \\
        \hline
        Spectroscopic Factors          & 0.632      & 0.703         & 0.422      \\
        \gmr Branching Ratios:\\
        \hspace*{1em} $> \val{6}{MeV}$ from $p$-hole & 0\%        & 91.8\%        & 14.7\%     \\
        \hspace*{1em} $> \val{6}{MeV}$ from $n$-hole & 0\%        & 86.9\%        & 14.7\%     \\
        \hspace*{1em} \val{3-6}{MeV} from either     & 0\%        & 0\%           & 27.8\%     \\
        \hline\hline
    \end{tabular}
    \caption{The spectroscopic factors and branching ratios for primary \gmr production for the three residual excited nuclear states simulated in this analysis. The spectroscopic factors are calculated in~\cite{Ankowski:2011ei}, while the branching ratios were measured with electron and proton scattering~\cite{Leuschner:1994zz,AjzenbergSelove:1991zz,Kobayashi:2006gb}.
    }
    \label{tab:specfac}
\end{table}

Neutrino interactions based on the above flux are simulated using the NEUT event generator~\cite{Nakahata:1986zp,Hayato:2002sd}. The NCQE cross section on oxygen is simulated using a spectral function model~\cite{Benhar:2005dj,Benhar:1994hw} with the BBBA05 form factor parameterization~\cite{Bradford:2006yz}, which is then reweighted as a function of neutrino energy to match the recent theoretical calculations from~\cite{Ankowski:2011ei}. In order to simulate the deexcitation \gmr emission, it is necessary to identify which state the remaining nucleus is in after the neutrino interaction. In the simple shell model, the nucleons in \atom{16}{O} occupy three states: $1p_{1/2}$, $1p_{3/2}$ and $1s_{1/2}$ with knockout energies of 12.1, 18.4, and ~42 MeV (3.54 MeV more for neutrons), though precise electron scattering measurements have shown significant deviations from this mean-field scenario~\cite{Bernheim:1981si,Leuschner:1994zz,Fissum:2004we}.  Approximately 20\% of the total spectral strength is pushed by $NN$ correlations out of the Fermi sea into continuum states~\cite{Rohe:2004dz}.  The electron scattering results are incorporated into a realistic spectral function model in~\cite{Benhar:2005dj}, which was integrated in momentum and energy in~\cite{Ankowski:2011ei} to calculate the spectroscopic factors for the shell states shown in \cref{tab:specfac}. 

Since the $1p_{1/2}$ hole is already the ground state, it produces no primary \gmrs.  The $1p_{3/2}$ proton and neutron hole states have three possible energy levels, the most common of which (branching ratio 87\%) has the lowest energy and decays by releasing a \val{6.32}{MeV} or \val{6.18}{MeV} photon for a proton or neutron hole, respectively.  That is the only $1p_{3/2}$ neutron-hole decay that produces a photon, but for the proton holes one of the higher energy levels also releases a \val{9.93}{MeV} photon, bringing the total branching ratio from \gmr production up to 92\%~\cite{Leuschner:1994zz,AjzenbergSelove:1991zz}.  
The $1s_{1/2}$ hole state can decay via a variety of channels, usually including additional nucleon emission because of the large binding energy of the knocked-out nucleon. The branching ratios of the $1s_{1/2}$ proton hole state are estimated using the result of the $\atom{16}{O}(p,2p)\atom{15}{N}$ experiment (RCNP-E148)~\cite{Kobayashi:2006gb}, with a \val{3-6}{MeV} \gmr produced 22\% of the time and a $>\val{6}{MeV}$ \gmr produced 15\% of the time.  Only protons were studied in the experiment, but the proton and neutron energy levels are expected to be very similar since \atom{16}{O} is an isoscalar nucleus.  There have been no measurements of photon production from the continuum states so they are assumed to produce no primary \gmrs, though this assumption is taken into account in the systematic uncertainties in \cref{sec:systematics}.

Non-QE NC interactions make up the largest neutrino-induced background component and predominantly consist of NC single-pion production where the pion is absorbed during final state interactions in the nucleus. This resonant production is simulated using the Rein-Sehgal model~\cite{Rein:1980wg}, the position dependence within the nucleus is calculated with the model from \cite{Oset:1982}, and the scale of the microscopic pion interaction probabilities in the nuclear medium is determined from fits to pion scattering data~\cite{Ashery:1981tq,Jones:1993ps,Giannelli:2000zy}.  The simulation of primary deexcitation \gmrs from this process is based on measurements of $\pi^-$ absorption-at-rest on H$_{2}$O at CERN~\cite{Engelhardt:1975ct}. These pion-absorption interactions can also release nucleons which go on to produce secondary \gmrs as described below.  More details about NEUT, including the models used to simulate the smaller charged current backgrounds can be found in~\cite{Abe:2011ks,Hayato:2002sd}.

SK's GEANT3-based simulation~\cite{Brun:1994aa} is used to transport all the particles leaving the nucleus through the detector, produce and transport the Cherenkov light, and to simulate the response of the photodetectors and electronics. Charged pions with momenta above \val{500}{MeV/\cc} are simulated with GCALOR~\cite{Zeitnitz:1994bs} while lower momentum pions are simulated with a custom routine based on the NEUT cascade model for final state hadrons.
GCALOR also simulates the interactions of nucleons with nuclei in the water, including the production of secondary \gmrs. 
In this simulation, secondary \gmrs are typically produced in multiples: 95\% of events with secondary \gmrs have at least two, some originating from multiple interactions of neutrons in the water. The total secondary \gmr energy per event is distributed widely with a peak around \val{7}{MeV} and a long tail towards higher energies. 

There is an additional signal-like contribution from the inelastic process, $\nu + \atom{16}{O}\rightarrow \nu + \atome{16}{O}$. However, since there is no accurate estimation of \gmr production induced by the NC inelastic process in the T2K energy range, we do not subtract its contribution in the final result. If we assume that the rate of \gmr production after a inelastic interaction is similar to that of a nucleon knockout reaction, and extrapolate the NC inelastic cross section predicted in~\cite{Kolbe:2002gk} to the energy region of this analysis, we expect its contribution to be no larger than a few percent of our final sample.

\section{Analysis}

The results presented in this paper are based on T2K RUN1-3 data from $3.01 \times 10^{20}$ protons on target (POT)~\cite{Abe:2013fuq}. 
The expected number of beam-related events after the selections described in the next section are summarized in Tab.~\ref{tab:mc_break}, which categorizes them by neutrino flavor  and interaction mode. For the computation of the CC components, we assume three-flavor oscillations with $|\Delta m_{32}^{2}|=2.44 \times 10^{-3}$ eV$^{2}$, $\sin^{2} \theta_{23}=0.50$, $\sin^{2} 2\theta_{13}=0.097$. The majority of the beam-related background comes from nonquasielastic NC events, in particular single-pion production followed by pion absorption within the nucleus. The CC background comes from interactions where the outgoing charged lepton has low momentum and is misidentified as an electron or where the charged lepton itself is below Cherenkov threshold but deexcitation \gmrs are emitted.

The expected number of beam-unrelated events after all selections are applied is estimated to be \bkgdnon by sampling events at least \val{5}{\mu s} before the T2K beam trigger so that no beam-related activity is included. The measured event rate is normalized to the total livetime of the analyzed beam spills. Since the beam-unrelated background is directly measured with data outside the beam window, the systematic uncertainty associated with it is small.

\begin{table}[b]
\caption{Observed and expected numbers of events in T2K Runs 1-3. The CC samples are based on the flux at SK including three-flavor oscillations (parameters described in the text). The NC samples are based on the unoscillated flux. The $\nue$ NCQE events are treated as signal, but the $\numubar$ NCQE are considered background since there is a different predicted cross-section for antineutrinos.} 
\begin{center}
\scalebox{0.88}{
\begin{tabular}{lc@{\tsp}c@{\tsp}c}
\hline\hline
Beam-related expectation & $\numu$ & $\nue$ & $\numubar$  \\
\hline
NCQE      &  34.33 &   0.46 &   0.69 \\
\ncoth    &  11.59 &   0.26 &   0.45 \\
CC        &   2.01 & 0.0014 &  0.025 \\
\hline
Signal           & \multicolumn{3}{c}{34.80} \\
Background (beam)& \multicolumn{3}{c}{15.02} \\
Beam-unrelated   & \multicolumn{3}{c}{ 1.20} \\
\hline
Observed events  & \multicolumn{3}{c}{\obs} \\
\hline\hline
\end{tabular}
\label{tab:mc_break}
}
\end{center}
\end{table}

\subsection{Event selection}

The event trigger for this analysis requires events to be within the \val{1}{ms} T2K beam window and have at least 25 PMT hits within \val{200}{ns}.  This trigger has an estimated efficiency or 99.5\% for \val{4}{MeV} events and is the lowest threshold trigger used in any T2K or Super-K analysis.  It is only possible because of the sharp reduction in accidental backgrounds due to the beam timing requirement. 

The reconstruction of the event vertex, direction, and energy is the same as that used in the SK solar neutrino analysis~\cite{Abe:2010hy}. The event vertex is found by a maximum likelihood fit to the timing residuals of the Cherenkov light, accounting for the dark noise rate \cite{Smy:icrc}.  The vertex resolution is approximately \val{125}{cm} at \val{4}{MeV} and improves to below \val{50}{cm} above \val{12}{MeV}.  The event direction is reconstructed by comparing the observed hit pattern with the MC expectation for a single electron ring using a likelihood function.  The reconstructed energy is defined as the total energy of a single electron that would have produced all Cherenkov photons in the event. It is calculated using the effective number of hit PMTs $N_{\rm{eff}}$, which corrects for the rate of multiple hits on a single PMT, scattered and reflected light, the water quality, the dark noise rate, the photocathode coverage as a function of time and position, and the PMT quantum efficiency.  The relationship between $N_{\rm{eff}}$ and energy is determined using a MC simulation of monoenergetic electrons. We use this definition because it is used by the SK low-energy reconstruction tools, though we know many events have multiple particles and a variety of particle species.  The vertex resolution, water quality, and energy scale are calibrated using a variety of sources, described in detail in \cite{Abe:2013gga}.

The first selections applied are a cut on the reconstructed energy, only allowing events between \val{4}{MeV} and \val{30}{MeV}, a standard fiducial volume cut of \val{2}{m} from the detector wall, and a tighter event timing cut. The neutrino beam spill has a bunch structure, reflecting the underlying proton bunch structure, with 6 or 8 bunches separated by \val{581}{ns} gaps, delivered every ~\val{3}{s}. The \val{\pm100}{ns} timing cut contains the neutrino beam bunch width which has an observed RMS of \val{24}{ns} at SK.  The time is synchronized between the near and far sites using a common-view GPS system and the bunch timing is calibrated using the higher energy T2K neutrino events at SK.

\begin{figure}
  \begin{center}
   \includegraphics[width=7cm]{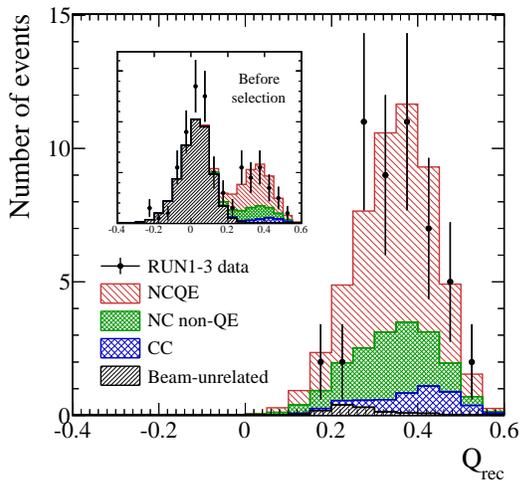}
  \end{center}
  \caption{(color online) Distribution of the reconstruction quality parameter, \q, after the beam-unrelated selection cuts (timing, fiducial volume, \q) have been applied. The inset shows the distribution before the energy-dependent cut on \q, but including the timing and fiducial volume cuts. The T2K RUN1-3 data are represented by points with statistical error bars and the expectation is represented by stacked histograms showing the NCQE signal and the \ncoth, CC, and beam-unrelated background components.}
  \label{fig:ovaq}
\end{figure}

\begin{figure}
  \begin{center}
   \includegraphics[width=7cm]{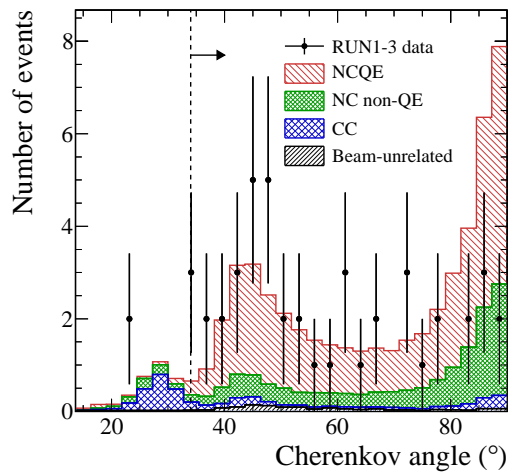}
  \end{center}
  \caption{(color online) The Cherenkov angle distribution in data and MC expectation after the beam-unrelated selections and the pre-activity cut. The expectation has a three-peak structure corresponding to low-momentum muons around 28\degr, single \gmrs around 42\degr, and multiple \gmrs around 90\degr. A selection cut is applied at 34\degr to remove the muon events, but no attempt is made to separate single- and multiple-$\gamma$ events.}
  \label{fig:changle}
\end{figure}

Further selection cuts are applied based on the event vertex and reconstruction quality to remove beam-unrelated background, similar to those used in SK solar~\cite{Abe:2010hy} and supernova relic neutrino analyses~\cite{Bays:2011si}. These cut criteria are simultaneously optimized in an energy-dependent way to maximize the figure-of-merit defined as $N_{\mathrm{beam}}/\sqrt{N_{\mathrm{beam}}+N_{\mathrm{unrel}}}$, where $N_{\mathrm{beam}}$ and $N_{\mathrm{unrel}}$ denote the number of expected beam-related and beam-unrelated events, respectively. The cut optimization is done separately for each of the three T2K run periods since the beam intensities and beam bunch structures differ. 

Most of the beam-unrelated background comes from radioactive impurities in the PMT glass, cases, and support structure and so is concentrated near the ID wall. Cuts on the distance from the nearest wall, \done, and the distance from the wall along the backward direction of the reconstructed track, \dtwo, together effectively eliminate background events produced at or near the ID wall. A minimum cut of \val{2}{m} is applied for both, with the cut on \done increasing linearly below \val{4.75}{MeV} to about \val{3.2}{m} and the cut on \dtwo increasing linearly below \val{5.75}{MeV} to about \val{10}{m}.

Beam-unrelated background events that pass the fiducial cuts typically have reconstruction errors which move the vertex to the center of the tank. These errors can be identified based on the distribution of hits in time and space. The hit time distribution should be a sharp peak after time-of-flight correction from the correct vertex, which we quantify as the timing goodness, \gt. The hit pattern should also be azimuthally symmetric around the reconstructed particle direction, which we test using \gp, the Kolmogorov-Smirnov distance between the observed hit distribution and a perfectly symmetric one. The reconstruction quality cut criterium, \q, is defined as the hyperbolic combination of these two parameters: $\q \equiv \gt^2 - \gp^2$ and is shown in Fig.~\ref{fig:ovaq}. The cut on \q is also energy-dependent and varies from about 0.25 at its tightest at the low end of the energy spectrum to effectively no cut above \val{11}{MeV}. More detailed descriptions of \gt and \gp are found in Ref.~\cite{Cravens:2008aa}.

Before selection, the beam-unrelated background rate from natural radioactivity is 284 counts per second, or 1.2 million events expected during the \val{1}{ms} beam windows used for other T2K analyses \cite{Abe:2011sj}. Applying the tight timing cut reduces this background to 1,816 events.  The fiducial and reconstruction quality cuts further reduce the beam-unrelated background to 1.77 events, a contamination of 2.4\%. These beam-unrelated selection cuts reduce the estimated NCQE signal efficiency to 74\%. Among the selected signal events, we estimate 97\% have true vertices within the fiducial volume.

Finally, to suppress the beam-related charged-current (CC) interaction events, two additional cuts are applied: a pre-activity cut and a Cherenkov opening angle cut. The pre-activity cut rejects electrons produced in muon decays with more than 99.9\% efficiency by rejecting events which occur less than \val{20}{\mu s} after a high-energy event, defined as a group of 22 or more hits in a \val{30}{ns} window.
The likelihood of this selection rejecting a signal event because of accidental dark noise hits is less than 0.1\%. 

For this low-energy sample, the Cherenkov angle of an event is defined as the peak of the distribution of Cherenkov angles calculated for every combination of three PMTs with hits, assuming the reconstructed vertex as the origin, following the technique from~~\cite{Bays:2011si}. For a single particle this peak will be close to the opening angle of the particle while the more isotropic light distributions from multiple particles will have peaks close to 90\degr.    
The Cherenkov angle depends on the velocity of the particle, approaching 42\degr as the velocity approaches $c$. The electrons produced by the deexcitation \gmrs selected in this analysis are highly relativistic and so peak at 42\degr. The heavier muons from $\numu$ CC events have smaller opening angles, peaking around 28\degr; the higher momentum muons with larger opening angles having already been removed by the energy cut at \val{30}{MeV}. These low-momentum muons are removed by a cut at 34\degr. The Cherenkov angle distribution for events passing all other selection criteria can be seen in Fig.~\ref{fig:changle}. The data-expectation disagreement in the multi-$\gamma$ peak is likely due to the approximations made in the model of \gmr emission induced by secondary neutron interactions used by GEANT3 and GCALOR, particularly the multiplicity of the secondary neutrons.

After all selections, $\mctot$ events are expected, of which $\ncel$ are expected to be NCQE signal for a purity of 69\%. The overall selection efficiency is estimated to be 70\% relative to the number of true NCQE events in the true fiducial volume which produce either primary or secondary $\gamma$-rays (approximately 25\% of NCQE events produce no photons and are consequently unobservable).  The beam-unrelated contamination remains 2.4\% after the final beam-related selections, with the 1.77 events after only the beam-unrelated selections reduced to \bkgdnon background events in the final sample.

\subsection{Observed Events}

\begin{figure}
  \begin{center}
   \includegraphics[width=7cm]{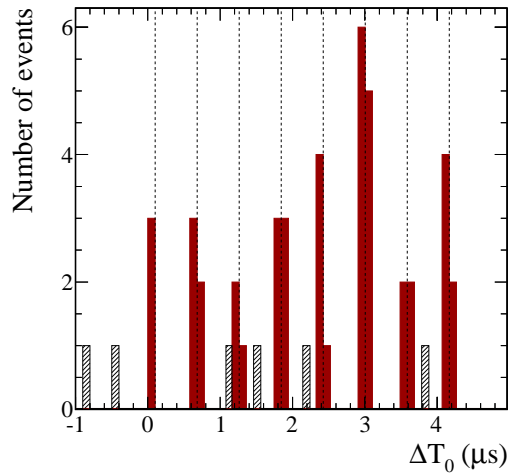}
  \end{center}
  \caption{(color online) $\Delta \mathrm{T}_{0}$ distribution of the data sample after all selection cuts except for beam timing, compared to the bunch center positions determined from high energy T2K neutrino events, indicated by eight dashed vertical lines. The on-timing and off-timing events are shown in solid and hashed, respectively.}
  \label{fig:dt0}
\end{figure}

Figure~\ref{fig:dt0} shows the observed event timing distribution in a region from \val{-1}{\mu s} to \val{5}{\mu s} with respect to the beam trigger time, before the tight \val{\pm100}{ns} timing cut on each bunch has been applied. Six events are found outside the tight bunch time windows, which is consistent with the 3.6 beam-unrelated events expected for this amount of integrated livetime. These events are separate from the \bkgdnon beam-unrelated events expected to fall within the \val{200}{ns} bunch windows.

\begin{figure}
  \begin{center}
   \includegraphics[width=7cm]{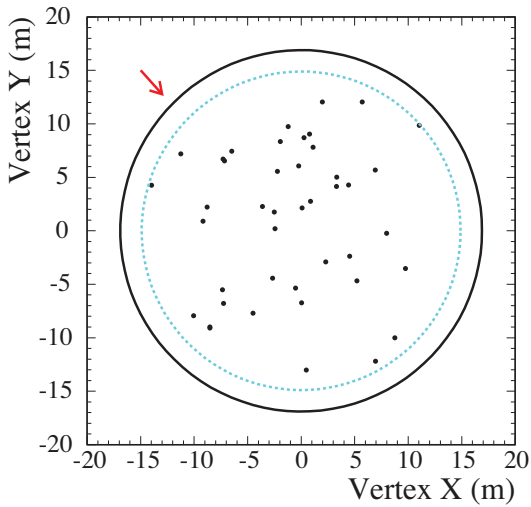}
  \end{center}
  \caption{(color online) Vertex distribution of the final data sample in Y vs X after all selections have been applied. The solid and dotted lines indicate the boundaries of the inner detector and fiducial volume, respectively. The neutrino beam direction is indicated by the arrow.}
  \label{fig:vtx}
\end{figure}

\begin{figure}
  \begin{center}
  \includegraphics[width=7cm]{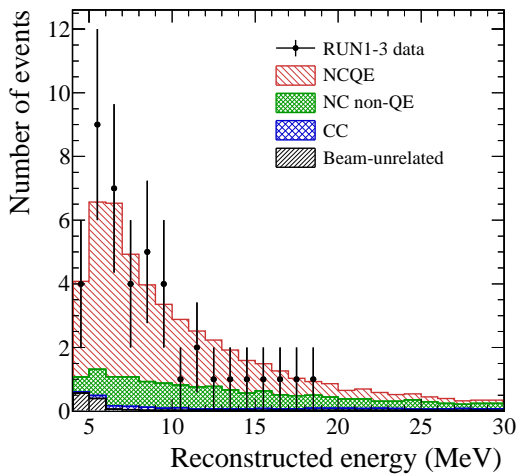}
  \end{center}
  \caption{(color online) Comparison of the reconstructed energy spectrum between the selected data sample and the expectation after all selection cuts have been applied.  The CC component is actually about twice as large as the beam-unrelated background, but it is less apparent since it is spread across all energies.}
  \label{fig:ene_comp}
\end{figure}

After all cuts, $\obs$ events remain in the \val{4-30}{MeV} reconstructed energy range, compared with $\mctot$ expected. The vertex distribution of the sample is shown in Fig.~\ref{fig:vtx}, in which no non-uniformity or biases with respect to the neutrino beam direction are found.  
The energy distribution of the data after all the selection cuts is shown in Fig.~\ref{fig:ene_comp}. A peak due to \val{6}{MeV} prompt deexcitation \gmrs is clearly seen in data, and the observed distribution matches well with the expectation. The high energy tail originates primarily from the contribution of additional secondary \gmrs overlapping the primary \gmrs.  

\subsection{Systematic uncertainties}\label{sec:systematics}

The sources of systematic uncertainty on the expected number of signal and background events and their size are summarized in Tab.~\ref{tab:sys_sum}. The methods for calculating these uncertainties are described below. 

The flux errors, calculated in correlated energy bins, are determined based on beam monitoring, constraints from external measurements (particularly NA61/SHINE~\cite{Abgrall:2011ae,Abgrall:2011ts}), and Monte Carlo studies of focusing parameters (e.g. horn current, beam alignment, etc.)~\cite{Abe:2012av}. The dominant uncertainty on the flux comes from the production of hadrons in the target, though the beam alignment and off-axis angle become significant at energies beween \val{700}{MeV} and \val{1.2}{GeV}, the high-energy side of the focusing peak. The neutrino interaction uncertainties which affect the normalization of the background are evaluated by comparing NEUT predictions to external neutrino-nucleus data sets in an energy region similar to T2K~\cite{Abe:2013jth}.

The 15\% systematic uncertainty on primary \gmr production in signal (and the QE component of the CC background) comes from several sources. The largest contribution comes from the uncertainty on the $1p_{3/2}$ spectroscopic factor.  We take the 34\% difference between the value used in our simulation (from~\cite{Ankowski:2011ei}) and the value calculated by Ejiri, 0.940~\cite{Ejiri:1993rh}, which gives an 11\% change in the number of selected NCQE events.  The second-largest contribution is from final-state nuclear interactions: NEUT assumes that the deexcitation \gmr production is the same whether the final state contains a single nucleon or multiple nucleons.  We estimate the systematic uncertainty introduced by this assumption by observing the change in the number of signal events with the extreme alternate assumption that no deexcitation \gmrs are released from events with multi-nucleon final states, which gives a 9\% change in the number of selected NCQE. 
Both of these errors have a diluted effect on the final sample since NCQE events can be observed via either primary or secondary \gmrs.

The $1s_{1/2}$ spectral function has a 30\% uncertainty, also calculated by model comparisons between~\cite{Ankowski:2011ei} and~\cite{Ejiri:1993rh}, but only gives a 1\% contribution to the final error.  The uncertainty in the $1s_{1/2}$ branching ratios is also accounted for based on the measurement from Kobayashi et al.~\cite{Kobayashi:2006gb}, giving an additional 1\% error.  The uncertainty due to not simulating photon production in the $NN$-induced continuum states is estimated by assuming these events emit photons in the same way as the $1s_{1/2}$ state, a reasonable approximation since both have binding energies well above the particle emission threshold, giving a 3\% error.  These systematics have only a small effect on the final selected sample since only a small fraction of $1s_{1/2}$ deexcitations produce photons above \val{6}{MeV}. 

For the non-QE NC background events, a conservative uncertainty was calculated by removing all primary \gmrs from the events and evaluating the difference in total selected events. The effect is relatively small since the pion-absorption events which make up the bulk of the \ncoth background produce many secondary \gmrs and so are still detected thanks to the low threshold of the analysis.

The uncertainty on secondary \gmr production is dominated by uncertainties on the production of neutrons.  It was evaluated by comparing alternate models of neutron production and interaction from GCALOR and NEUT and how they altered the observed Cherenkov light level for our simulated events, for both signal events and the pion-absorption background.  Even with changes of a factor of two or more in the average amount of energy deposited in the detector between the different neutron interaction models, the change in the number of selected events is relatively small because we only count events and do not attempt to distinguish between events with one or multiple \gmrs. Events with secondary \gmrs typically have multiple \gmrs and often have primary \gmrs, as well, so the detection efficiency is kept high even with a significant change in the secondary \gmr production probability.

The detector uncertainty includes contributions from uncertainties in the SK energy scale, vertex resolution, and selection efficiency.  It is estimated by comparing simulation and data from the linear electron accelerator (LINAC) installed above SK~\cite{Nakahata:1998pz}. The systematic uncertainty due to the atmospheric oscillation parameters, $\theta_{23}$ and $|\Delta m^{2}_{32}|$, is estimated by varying the parameters within their uncertainties from the T2K measurement of these parameters~\cite{Abe:2013fuq}.

There are two final systematic uncertainties that were evaluated but have a negligible impact on the result. We evaluated the potential non-uniformity of the selection efficiency with respect to $Q^2$ by changing the value of the MC axial mass to distort the differential cross section. This variation changes the final calculated cross section by less than a percent.
The beam-unrelated background is estimated from the out-of-time events which have a statistical error of $0.8\%$.

\begin{table}
\caption{Summary of systematic uncertainties on the expected number of signal and background events. While the CC component has the largest uncertainty, it has a relatively small effect on the final result since there are relatively few CC events in the final sample.}
\label{tab:sys_sum}
\begin{center}
\begin{tabular}{lcccc}
\hline\hline
                              & Signal     & \multicolumn{3}{c}{Background}     \\
                              & NCQE       & \ncoth   & CC     & Unrel. \\
Fraction of Sample            & 68\%       & 26\%     &  4\%   & 2\%            \\
\hline
Flux                          & 11\%       & 10\%     & 12\%   & - \\
Cross sections                &  -         & 18\%     & 24\%   & - \\
Primary $\gamma$ production   & 15\%       &  3\%     &  9\%   & - \\
Secondary $\gamma$ production & 13\%       & 13\%     &  7.6\% & - \\
Detector response             &  2.2\%     &  2.2\%   &  2.2\% & - \\
Oscillation Parameters        &  -         &  -       & 10\%   & - \\
\hline
Total Systematic Error        & 23\%       & 25\%     & 31\%   & 0.8\% \\
\hline\hline
\end{tabular}
\end{center}
\end{table}

\section{Measured cross section}

\begin{figure}
 \begin{center}
  \includegraphics[width=7cm]{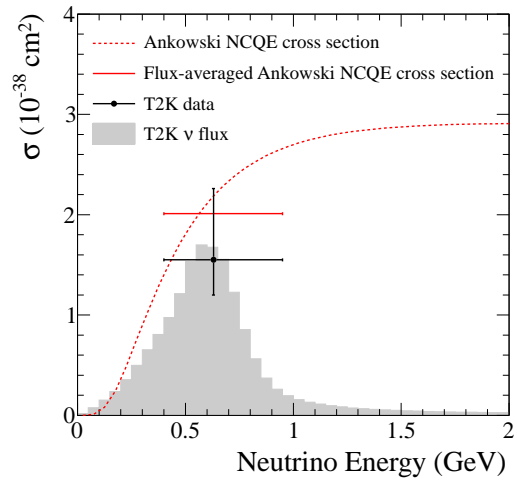}
  \caption{(color online) The T2K measurement of the flux-averaged NCQE cross section, represented by a black point, compared with the calculated cross section from~\cite{Ankowski:2011ei}. The dashed line shows the cross section versus neutrino energy, the solid horizontal line shows the flux-averaged cross section. The vertical error bar on the data represents the 68\% confidence interval on the measured cross section while the horizontal error bar is placed at the central value from our data and represents 68\% of the flux at each side of the \fluxtype energy. The solid gray histogram shows the unoscillated T2K neutrino flux.}
  \label{fig:ncqexsec}
 \end{center}
\end{figure}

The NCQE cross section is measured by comparing the NCQE cross section as calculated in recent theoretical work~\cite{Ankowski:2011ei} averaged over the unoscillated T2K flux with the observed number of events after background subtraction:
\begin{equation}\label{eq:xsec}
\langle \sigma_{\nu, \mathrm{NCQE}}^{obs} \rangle = \frac{N^{obs}-N_{bkg}^{exp}}{N^{exp}-N_{bkg}^{exp}}\langle \sigma_{\nu, \mathrm{NCQE}}^{theory} \rangle,
\end{equation}
where $\langle \sigma_{\nu, \mathrm{NCQE}}^{obs} \rangle$ is the observed flux-averaged NCQE cross section and $\langle \sigma_{\nu, \mathrm{NCQE}}^{theory} \rangle = \val{\xsecpre \times 10^{-38}}{cm^2}$ is the flux-averaged cross section from~\cite{Ankowski:2011ei}.  The NCQE cross section includes all spectral states since even states without primary \gmr emission are often observed with secondary \gmrs. The total number of observed events is $N^{obs}$~(\obs), the total number of expected events is $N^{exp}$~(\mctot), and $N_{bkg}^{exp}$~(\bkgd) denotes the expected number of background events. 

The obtained flux-averaged neutrino-oxygen NCQE cross section is \val{\xsec \times 10^{-38}}{cm^2} at a \fluxtype neutrino flux energy of \fluxmean. The 68\% confidence interval on the cross section is \val{(\xseclow, \xsechigh) \times 10^{-38}}{cm^2} and the 90\% confidence interval is \val{(\xseclowninety, \xsechighninety) \times 10^{-38}}{cm^2}. They include both statistical and systematic errors and were calculated using a Monte Carlo method to account for the systematic errors that are correlated between different samples.  
While the underlying systematic uncertainties are symmetric and gaussian, the confidence interval is asymmetric around the central value because some of the uncertainties, primarily the production of secondary $\gamma$-rays and to a lesser extent the neutrino flux, are correlated between the background expectation and the signal expectation which are found in the numerator and denominator, respectively, of Eq.~\ref{eq:xsec}. Figure~\ref{fig:ncqexsec} shows our result compared with a theoretical calculation of the NCQE cross section~\cite{Ankowski:2011ei}. The vertical error bar for data shows the 68\% confidence interval on the data, and the horizontal error bar represents 68\% of the flux at each side of the \fluxtype energy. The measurement is consistent with the recent theoretical calculation at the 68\% confidence level.  Using the same Monte Carlo method used to calculate the confidence intervals, we reject the background-only hypothesis (NCQE cross section of 0) with a $p$-value of $4 \times 10^{-8}$, which corresponds to a one-sided $z$-score of $5.4\sigma$.

\section{Conclusion}
We have reported the first measurement of the cross section of neutrino-oxygen NCQE interactions, measured via the detection of nuclear deexcitation \gmrs in the Super-Kamiokande detector using the T2K narrow-band neutrino beam.  Our measurement is consistent with the theoretical expectation at the 68\% confidence level. Due to the similar peak energies for T2K neutrinos and atmospheric neutrinos, the present work will shed light on the study of the atmospheric background events for low energy astrophysical  phenomena in neutrino experiments. In this paper we assume that oscillations do not modify the total flux of neutrinos at SK, but oscillations to a sterile neutrino (such as those considered in~\cite{Adamson:2011ku}) could lower the overall rate of all NC events, including NCQE events.

\section{Acknowledgements}
We thank the J-PARC staff for superb accelerator performance and the
CERN NA61 collaboration for providing valuable particle production data.
We acknowledge the support of MEXT, Japan;
NSERC, NRC and CFI, Canada;
CEA and CNRS/IN2P3, France;
DFG, Germany;
INFN, Italy;
National Science Centre (NCN), Poland;
RSF, RFBR and MES, Russia;
MINECO and ERDF funds, Spain;
SNSF and SER, Switzerland;
STFC, UK; and
DOE, USA.
We also thank CERN for the UA1/NOMAD magnet,
DESY for the HERA-B magnet mover system,
NII for SINET4,
the WestGrid and SciNet consortia in Compute Canada,
GridPP, UK.
In addition participation of individual researchers
and institutions has been further supported by funds from: ERC (FP7), EU;
JSPS, Japan;
Royal Society, UK;
DOE Early Career program, USA.


\bibliography{ncgamma}

\begin{thebibliography}{50}%
\makeatletter
\providecommand \@ifxundefined [1]{%
 \@ifx{#1\undefined}
}%
\providecommand \@ifnum [1]{%
 \ifnum #1\expandafter \@firstoftwo
 \else \expandafter \@secondoftwo
 \fi
}%
\providecommand \@ifx [1]{%
 \ifx #1\expandafter \@firstoftwo
 \else \expandafter \@secondoftwo
 \fi
}%
\providecommand \natexlab [1]{#1}%
\providecommand \enquote  [1]{``#1''}%
\providecommand \bibnamefont  [1]{#1}%
\providecommand \bibfnamefont [1]{#1}%
\providecommand \citenamefont [1]{#1}%
\providecommand \href@noop [0]{\@secondoftwo}%
\providecommand \href [0]{\begingroup \@sanitize@url \@href}%
\providecommand \@href[1]{\@@startlink{#1}\@@href}%
\providecommand \@@href[1]{\endgroup#1\@@endlink}%
\providecommand \@sanitize@url [0]{\catcode `\\12\catcode `\$12\catcode
  `\&12\catcode `\#12\catcode `\^12\catcode `\_12\catcode `\%12\relax}%
\providecommand \@@startlink[1]{}%
\providecommand \@@endlink[0]{}%
\providecommand \url  [0]{\begingroup\@sanitize@url \@url }%
\providecommand \@url [1]{\endgroup\@href {#1}{\urlprefix }}%
\providecommand \urlprefix  [0]{URL }%
\providecommand \Eprint [0]{\href }%
\providecommand \doibase [0]{http://dx.doi.org/}%
\providecommand \selectlanguage [0]{\@gobble}%
\providecommand \bibinfo  [0]{\@secondoftwo}%
\providecommand \bibfield  [0]{\@secondoftwo}%
\providecommand \translation [1]{[#1]}%
\providecommand \BibitemOpen [0]{}%
\providecommand \bibitemStop [0]{}%
\providecommand \bibitemNoStop [0]{.\EOS\space}%
\providecommand \EOS [0]{\spacefactor3000\relax}%
\providecommand \BibitemShut  [1]{\csname bibitem#1\endcsname}%
\let\auto@bib@innerbib\@empty
\bibitem [{\citenamefont {Bodmann}\ \emph {et~al.}(1992)\citenamefont {Bodmann}
  \emph {et~al.}}]{Bodmann:1992ur}%
  \BibitemOpen
  \bibfield  {author} {\bibinfo {author} {\bibfnamefont {B.}~\bibnamefont
  {Bodmann}} \emph {et~al.} (\bibinfo {collaboration} {KARMEN Collaboration}),\
  }\href {\doibase 10.1016/0370-2693(92)90055-9} {\bibfield  {journal}
  {\bibinfo  {journal} {Phys.Lett.}\ }\textbf {\bibinfo {volume} {B280}},\
  \bibinfo {pages} {198} (\bibinfo {year} {1992})}\BibitemShut {NoStop}%
\bibitem [{\citenamefont {Armbruster}\ \emph {et~al.}(1998)\citenamefont
  {Armbruster} \emph {et~al.}}]{Armbruster:1998gk}%
  \BibitemOpen
  \bibfield  {author} {\bibinfo {author} {\bibfnamefont {B.}~\bibnamefont
  {Armbruster}} \emph {et~al.} (\bibinfo {collaboration} {KARMEN
  Collaboration}),\ }\href {\doibase 10.1016/S0370-2693(98)00087-2} {\bibfield
  {journal} {\bibinfo  {journal} {Phys.Lett.}\ }\textbf {\bibinfo {volume}
  {B423}},\ \bibinfo {pages} {15} (\bibinfo {year} {1998})}\BibitemShut
  {NoStop}%
\bibitem [{\citenamefont {Langanke}\ \emph {et~al.}(1996)\citenamefont
  {Langanke}, \citenamefont {Vogel},\ and\ \citenamefont
  {Kolbe}}]{Langanke:1995he}%
  \BibitemOpen
  \bibfield  {author} {\bibinfo {author} {\bibfnamefont {K.}~\bibnamefont
  {Langanke}}, \bibinfo {author} {\bibfnamefont {P.}~\bibnamefont {Vogel}}, \
  and\ \bibinfo {author} {\bibfnamefont {E.}~\bibnamefont {Kolbe}},\ }\href
  {\doibase 10.1103/PhysRevLett.76.2629} {\bibfield  {journal} {\bibinfo
  {journal} {Phys.Rev.Lett.}\ }\textbf {\bibinfo {volume} {76}},\ \bibinfo
  {pages} {2629} (\bibinfo {year} {1996})},\ \Eprint
  {http://arxiv.org/abs/nucl-th/9511032} {arXiv:nucl-th/9511032 [nucl-th]}
  \BibitemShut {NoStop}%
\bibitem [{\citenamefont {Kolbe}\ \emph {et~al.}(1992)\citenamefont {Kolbe},
  \citenamefont {Langanke}, \citenamefont {Krewald},\ and\ \citenamefont
  {Thielemann}}]{Kolbe:1992xu}%
  \BibitemOpen
  \bibfield  {author} {\bibinfo {author} {\bibfnamefont {E.}~\bibnamefont
  {Kolbe}}, \bibinfo {author} {\bibfnamefont {K.}~\bibnamefont {Langanke}},
  \bibinfo {author} {\bibfnamefont {S.}~\bibnamefont {Krewald}}, \ and\
  \bibinfo {author} {\bibfnamefont {F.}~\bibnamefont {Thielemann}},\ }\href
  {\doibase 10.1016/0375-9474(92)90175-J} {\bibfield  {journal} {\bibinfo
  {journal} {Nucl.Phys.}\ }\textbf {\bibinfo {volume} {A540}},\ \bibinfo
  {pages} {599} (\bibinfo {year} {1992})}\BibitemShut {NoStop}%
\bibitem [{\citenamefont {Beacom}\ and\ \citenamefont
  {Vogel}(1998{\natexlab{a}})}]{Beacom:1998ya}%
  \BibitemOpen
  \bibfield  {author} {\bibinfo {author} {\bibfnamefont {J.~F.}\ \bibnamefont
  {Beacom}}\ and\ \bibinfo {author} {\bibfnamefont {P.}~\bibnamefont {Vogel}},\
  }\href {\doibase 10.1103/PhysRevD.58.053010} {\bibfield  {journal} {\bibinfo
  {journal} {Phys.Rev.}\ }\textbf {\bibinfo {volume} {D58}},\ \bibinfo {pages}
  {053010} (\bibinfo {year} {1998}{\natexlab{a}})},\ \Eprint
  {http://arxiv.org/abs/hep-ph/9802424} {arXiv:hep-ph/9802424 [hep-ph]}
  \BibitemShut {NoStop}%
\bibitem [{\citenamefont {Beacom}\ and\ \citenamefont
  {Vogel}(1998{\natexlab{b}})}]{Beacom:1998yb}%
  \BibitemOpen
  \bibfield  {author} {\bibinfo {author} {\bibfnamefont {J.~F.}\ \bibnamefont
  {Beacom}}\ and\ \bibinfo {author} {\bibfnamefont {P.}~\bibnamefont {Vogel}},\
  }\href {\doibase 10.1103/PhysRevD.58.093012} {\bibfield  {journal} {\bibinfo
  {journal} {Phys.Rev.}\ }\textbf {\bibinfo {volume} {D58}},\ \bibinfo {pages}
  {093012} (\bibinfo {year} {1998}{\natexlab{b}})},\ \Eprint
  {http://arxiv.org/abs/hep-ph/9806311} {arXiv:hep-ph/9806311 [hep-ph]}
  \BibitemShut {NoStop}%
\bibitem [{\citenamefont {Nussinov}\ and\ \citenamefont
  {Shrock}(2001)}]{Nussinov:2000qc}%
  \BibitemOpen
  \bibfield  {author} {\bibinfo {author} {\bibfnamefont {S.}~\bibnamefont
  {Nussinov}}\ and\ \bibinfo {author} {\bibfnamefont {R.}~\bibnamefont
  {Shrock}},\ }\href {\doibase 10.1103/PhysRevLett.86.2223} {\bibfield
  {journal} {\bibinfo  {journal} {Phys.Rev.Lett.}\ }\textbf {\bibinfo {volume}
  {86}},\ \bibinfo {pages} {2223} (\bibinfo {year} {2001})},\ \Eprint
  {http://arxiv.org/abs/hep-ph/0009334} {arXiv:hep-ph/0009334 [hep-ph]}
  \BibitemShut {NoStop}%
\bibitem [{\citenamefont {Kolbe}\ \emph {et~al.}(2002)\citenamefont {Kolbe},
  \citenamefont {Langanke},\ and\ \citenamefont {Vogel}}]{Kolbe:2002gk}%
  \BibitemOpen
  \bibfield  {author} {\bibinfo {author} {\bibfnamefont {E.}~\bibnamefont
  {Kolbe}}, \bibinfo {author} {\bibfnamefont {K.}~\bibnamefont {Langanke}}, \
  and\ \bibinfo {author} {\bibfnamefont {P.}~\bibnamefont {Vogel}},\ }\href
  {\doibase 10.1103/PhysRevD.66.013007} {\bibfield  {journal} {\bibinfo
  {journal} {Phys.Rev.}\ }\textbf {\bibinfo {volume} {D66}},\ \bibinfo {pages}
  {013007} (\bibinfo {year} {2002})}\BibitemShut {NoStop}%
\bibitem [{\citenamefont {Ankowski}\ \emph {et~al.}(2012)\citenamefont
  {Ankowski}, \citenamefont {Benhar}, \citenamefont {Mori}, \citenamefont
  {Yamaguchi},\ and\ \citenamefont {Sakuda}}]{Ankowski:2011ei}%
  \BibitemOpen
  \bibfield  {author} {\bibinfo {author} {\bibfnamefont {A.~M.}\ \bibnamefont
  {Ankowski}}, \bibinfo {author} {\bibfnamefont {O.}~\bibnamefont {Benhar}},
  \bibinfo {author} {\bibfnamefont {T.}~\bibnamefont {Mori}}, \bibinfo {author}
  {\bibfnamefont {R.}~\bibnamefont {Yamaguchi}}, \ and\ \bibinfo {author}
  {\bibfnamefont {M.}~\bibnamefont {Sakuda}},\ }\href {\doibase
  10.1103/PhysRevLett.108.052505} {\bibfield  {journal} {\bibinfo  {journal}
  {Phys.Rev.Lett.}\ }\textbf {\bibinfo {volume} {108}},\ \bibinfo {pages}
  {052505} (\bibinfo {year} {2012})},\ \Eprint {http://arxiv.org/abs/1110.0679}
  {arXiv:1110.0679 [nucl-th]} \BibitemShut {NoStop}%
\bibitem [{\citenamefont {Bays}\ \emph {et~al.}(2012)\citenamefont {Bays} \emph
  {et~al.}}]{Bays:2011si}%
  \BibitemOpen
  \bibfield  {author} {\bibinfo {author} {\bibfnamefont {K.}~\bibnamefont
  {Bays}} \emph {et~al.} (\bibinfo {collaboration} {Super-Kamiokande
  Collaboration}),\ }\href {\doibase 10.1103/PhysRevD.85.052007} {\bibfield
  {journal} {\bibinfo  {journal} {Phys.Rev.}\ }\textbf {\bibinfo {volume}
  {D85}},\ \bibinfo {pages} {052007} (\bibinfo {year} {2012})},\ \Eprint
  {http://arxiv.org/abs/1111.5031} {arXiv:1111.5031 [hep-ex]} \BibitemShut
  {NoStop}%
\bibitem [{\citenamefont {Ueno}\ \emph {et~al.}(2012)\citenamefont {Ueno} \emph
  {et~al.}}]{Ueno:2012md}%
  \BibitemOpen
  \bibfield  {author} {\bibinfo {author} {\bibfnamefont {K.}~\bibnamefont
  {Ueno}} \emph {et~al.} (\bibinfo {collaboration} {Super-Kamiokande
  Collaboration}),\ }\href {\doibase 10.1016/j.astropartphys.2012.05.008}
  {\bibfield  {journal} {\bibinfo  {journal} {Astropart.Phys.}\ }\textbf
  {\bibinfo {volume} {36}},\ \bibinfo {pages} {131} (\bibinfo {year} {2012})},\
  \Eprint {http://arxiv.org/abs/1203.0940} {arXiv:1203.0940 [hep-ex]}
  \BibitemShut {NoStop}%
\bibitem [{\citenamefont {Fukuda}\ \emph {et~al.}(2003)\citenamefont {Fukuda}
  \emph {et~al.}}]{Fukuda:2002uc}%
  \BibitemOpen
  \bibfield  {author} {\bibinfo {author} {\bibfnamefont {Y.}~\bibnamefont
  {Fukuda}} \emph {et~al.} (\bibinfo {collaboration} {Super-Kamiokande
  Collaboration}),\ }\href {\doibase 10.1016/S0168-9002(03)00425-X} {\bibfield
  {journal} {\bibinfo  {journal} {Nucl.Instrum.Meth.}\ }\textbf {\bibinfo
  {volume} {A501}},\ \bibinfo {pages} {418} (\bibinfo {year}
  {2003})}\BibitemShut {NoStop}%
\bibitem [{\citenamefont {Beacom}\ and\ \citenamefont
  {Vagins}(2004)}]{Beacom:2003nk}%
  \BibitemOpen
  \bibfield  {author} {\bibinfo {author} {\bibfnamefont {J.~F.}\ \bibnamefont
  {Beacom}}\ and\ \bibinfo {author} {\bibfnamefont {M.~R.}\ \bibnamefont
  {Vagins}},\ }\href {\doibase 10.1103/PhysRevLett.93.171101} {\bibfield
  {journal} {\bibinfo  {journal} {Phys.Rev.Lett.}\ }\textbf {\bibinfo {volume}
  {93}},\ \bibinfo {pages} {171101} (\bibinfo {year} {2004})},\ \Eprint
  {http://arxiv.org/abs/hep-ph/0309300} {arXiv:hep-ph/0309300 [hep-ph]}
  \BibitemShut {NoStop}%
\bibitem [{\citenamefont {Abe}\ \emph {et~al.}(2011{\natexlab{a}})\citenamefont
  {Abe}, \citenamefont {Abe}, \citenamefont {Aihara}, \citenamefont {Fukuda},
  \citenamefont {Hayato} \emph {et~al.}}]{Abe:2011ts}%
  \BibitemOpen
  \bibfield  {author} {\bibinfo {author} {\bibfnamefont {K.}~\bibnamefont
  {Abe}}, \bibinfo {author} {\bibfnamefont {T.}~\bibnamefont {Abe}}, \bibinfo
  {author} {\bibfnamefont {H.}~\bibnamefont {Aihara}}, \bibinfo {author}
  {\bibfnamefont {Y.}~\bibnamefont {Fukuda}}, \bibinfo {author} {\bibfnamefont
  {Y.}~\bibnamefont {Hayato}},  \emph {et~al.},\ }\href@noop {} {\enquote
  {\bibinfo {title} {{Letter of Intent: The Hyper-Kamiokande Experiment ---
  Detector Design and Physics Potential ---}},}\ } (\bibinfo {year}
  {2011}{\natexlab{a}}),\ \Eprint {http://arxiv.org/abs/1109.3262}
  {arXiv:1109.3262 [hep-ex]} \BibitemShut {NoStop}%
\bibitem [{\citenamefont {Akiri}\ \emph {et~al.}(2011)\citenamefont {Akiri}
  \emph {et~al.}}]{Akiri:2011dv}%
  \BibitemOpen
  \bibfield  {author} {\bibinfo {author} {\bibfnamefont {T.}~\bibnamefont
  {Akiri}} \emph {et~al.} (\bibinfo {collaboration} {LBNE Collaboration}),\
  }\href@noop {} {\enquote {\bibinfo {title} {{The 2010 Interim Report of the
  Long-Baseline Neutrino Experiment Collaboration Physics Working Groups}},}\ }
  (\bibinfo {year} {2011}),\ \Eprint {http://arxiv.org/abs/1110.6249}
  {arXiv:1110.6249 [hep-ex]} \BibitemShut {NoStop}%
\bibitem [{\citenamefont {Abe}\ \emph {et~al.}(2011{\natexlab{b}})\citenamefont
  {Abe} \emph {et~al.}}]{Abe:2011ks}%
  \BibitemOpen
  \bibfield  {author} {\bibinfo {author} {\bibfnamefont {K.}~\bibnamefont
  {Abe}} \emph {et~al.} (\bibinfo {collaboration} {T2K Collaboration}),\ }\href
  {\doibase 10.1016/j.nima.2011.06.067} {\bibfield  {journal} {\bibinfo
  {journal} {Nucl.Instrum.Meth.}\ }\textbf {\bibinfo {volume} {A659}},\
  \bibinfo {pages} {106} (\bibinfo {year} {2011}{\natexlab{b}})},\ \Eprint
  {http://arxiv.org/abs/1106.1238} {arXiv:1106.1238 [physics.ins-det]}
  \BibitemShut {NoStop}%
\bibitem [{\citenamefont {Abe}\ \emph {et~al.}(2013{\natexlab{a}})\citenamefont
  {Abe} \emph {et~al.}}]{Abe:2013jth}%
  \BibitemOpen
  \bibfield  {author} {\bibinfo {author} {\bibfnamefont {K.}~\bibnamefont
  {Abe}} \emph {et~al.} (\bibinfo {collaboration} {T2K Collaboration}),\ }\href
  {\doibase 10.1103/PhysRevD.87.092003} {\bibfield  {journal} {\bibinfo
  {journal} {Phys.Rev.}\ }\textbf {\bibinfo {volume} {D87}},\ \bibinfo {pages}
  {092003} (\bibinfo {year} {2013}{\natexlab{a}})},\ \Eprint
  {http://arxiv.org/abs/1302.4908} {arXiv:1302.4908 [hep-ex]} \BibitemShut
  {NoStop}%
\bibitem [{\citenamefont {Battistoni}\ \emph {et~al.}(2011)\citenamefont
  {Battistoni}, \citenamefont {Broggi}, \citenamefont {Brugger}, \citenamefont
  {Campanella}, \citenamefont {Carboni} \emph {et~al.}}]{Battistoni:2011zz}%
  \BibitemOpen
  \bibfield  {author} {\bibinfo {author} {\bibfnamefont {G.}~\bibnamefont
  {Battistoni}}, \bibinfo {author} {\bibfnamefont {F.}~\bibnamefont {Broggi}},
  \bibinfo {author} {\bibfnamefont {M.}~\bibnamefont {Brugger}}, \bibinfo
  {author} {\bibfnamefont {M.}~\bibnamefont {Campanella}}, \bibinfo {author}
  {\bibfnamefont {M.}~\bibnamefont {Carboni}},  \emph {et~al.},\ }\href
  {\doibase 10.1016/j.nimb.2011.04.028} {\bibfield  {journal} {\bibinfo
  {journal} {Nucl.Instrum.Meth.}\ }\textbf {\bibinfo {volume} {B269}},\
  \bibinfo {pages} {2850} (\bibinfo {year} {2011})}\BibitemShut {NoStop}%
\bibitem [{\citenamefont {Abgrall}\ \emph {et~al.}(2011)\citenamefont {Abgrall}
  \emph {et~al.}}]{Abgrall:2011ae}%
  \BibitemOpen
  \bibfield  {author} {\bibinfo {author} {\bibfnamefont {N.}~\bibnamefont
  {Abgrall}} \emph {et~al.} (\bibinfo {collaboration} {NA61/SHINE
  Collaboration}),\ }\href {\doibase 10.1103/PhysRevC.84.034604} {\bibfield
  {journal} {\bibinfo  {journal} {Phys.Rev.}\ }\textbf {\bibinfo {volume}
  {C84}},\ \bibinfo {pages} {034604} (\bibinfo {year} {2011})},\ \Eprint
  {http://arxiv.org/abs/1102.0983} {arXiv:1102.0983 [hep-ex]} \BibitemShut
  {NoStop}%
\bibitem [{\citenamefont {Abgrall}\ \emph {et~al.}(2012)\citenamefont {Abgrall}
  \emph {et~al.}}]{Abgrall:2011ts}%
  \BibitemOpen
  \bibfield  {author} {\bibinfo {author} {\bibfnamefont {N.}~\bibnamefont
  {Abgrall}} \emph {et~al.} (\bibinfo {collaboration} {NA61/SHINE
  Collaboration}),\ }\href {\doibase 10.1103/PhysRevC.85.035210} {\bibfield
  {journal} {\bibinfo  {journal} {Phys.Rev.}\ }\textbf {\bibinfo {volume}
  {C85}},\ \bibinfo {pages} {035210} (\bibinfo {year} {2012})},\ \Eprint
  {http://arxiv.org/abs/1112.0150} {arXiv:1112.0150 [hep-ex]} \BibitemShut
  {NoStop}%
\bibitem [{\citenamefont {Bonesini}\ \emph {et~al.}(2001)\citenamefont
  {Bonesini}, \citenamefont {Marchionni}, \citenamefont {Pietropaolo},\ and\
  \citenamefont {Tabarelli~de Fatis}}]{Bonesini:2001iz}%
  \BibitemOpen
  \bibfield  {author} {\bibinfo {author} {\bibfnamefont {M.}~\bibnamefont
  {Bonesini}}, \bibinfo {author} {\bibfnamefont {A.}~\bibnamefont
  {Marchionni}}, \bibinfo {author} {\bibfnamefont {F.}~\bibnamefont
  {Pietropaolo}}, \ and\ \bibinfo {author} {\bibfnamefont {T.}~\bibnamefont
  {Tabarelli~de Fatis}},\ }\href {\doibase 10.1007/s100520100656} {\bibfield
  {journal} {\bibinfo  {journal} {Eur.Phys.J.}\ }\textbf {\bibinfo {volume}
  {C20}},\ \bibinfo {pages} {13} (\bibinfo {year} {2001})},\ \Eprint
  {http://arxiv.org/abs/hep-ph/0101163} {arXiv:hep-ph/0101163 [hep-ph]}
  \BibitemShut {NoStop}%
\bibitem [{\citenamefont {Brun}\ \emph {et~al.}(1994)\citenamefont {Brun},
  \citenamefont {Carminati},\ and\ \citenamefont {Giani}}]{Brun:1994aa}%
  \BibitemOpen
  \bibfield  {author} {\bibinfo {author} {\bibfnamefont {R.}~\bibnamefont
  {Brun}}, \bibinfo {author} {\bibfnamefont {F.}~\bibnamefont {Carminati}}, \
  and\ \bibinfo {author} {\bibfnamefont {S.}~\bibnamefont {Giani}},\
  }\href@noop {} {\enquote {\bibinfo {title} {{GEANT Detector Description and
  Simulation Tool}},}\ } (\bibinfo {year} {1994}),\ \bibinfo {note}
  {\mbox{CERN-W5013}}\BibitemShut {NoStop}%
\bibitem [{\citenamefont {Zeitnitz}\ and\ \citenamefont
  {Gabriel}(1994)}]{Zeitnitz:1994bs}%
  \BibitemOpen
  \bibfield  {author} {\bibinfo {author} {\bibfnamefont {C.}~\bibnamefont
  {Zeitnitz}}\ and\ \bibinfo {author} {\bibfnamefont {T.}~\bibnamefont
  {Gabriel}},\ }\href {\doibase 10.1016/0168-9002(94)90613-0} {\bibfield
  {journal} {\bibinfo  {journal} {Nucl.Instrum.Meth.}\ }\textbf {\bibinfo
  {volume} {A349}},\ \bibinfo {pages} {106} (\bibinfo {year}
  {1994})}\BibitemShut {NoStop}%
\bibitem [{\citenamefont {Abe}\ \emph {et~al.}(2013{\natexlab{b}})\citenamefont
  {Abe} \emph {et~al.}}]{Abe:2012av}%
  \BibitemOpen
  \bibfield  {author} {\bibinfo {author} {\bibfnamefont {K.}~\bibnamefont
  {Abe}} \emph {et~al.} (\bibinfo {collaboration} {T2K Collaboration}),\ }\href
  {\doibase 10.1103/PhysRevD.87.012001, 10.1103/PhysRevD.87.019902} {\bibfield
  {journal} {\bibinfo  {journal} {Phys.Rev.}\ }\textbf {\bibinfo {volume}
  {D87}},\ \bibinfo {pages} {012001} (\bibinfo {year} {2013}{\natexlab{b}})},\
  \Eprint {http://arxiv.org/abs/1211.0469} {arXiv:1211.0469 [hep-ex]}
  \BibitemShut {NoStop}%
\bibitem [{\citenamefont {Leuschner}\ \emph {et~al.}(1994)\citenamefont
  {Leuschner}, \citenamefont {Calarco}, \citenamefont {Hersman}, \citenamefont
  {Jans}, \citenamefont {Kramer} \emph {et~al.}}]{Leuschner:1994zz}%
  \BibitemOpen
  \bibfield  {author} {\bibinfo {author} {\bibfnamefont {M.}~\bibnamefont
  {Leuschner}}, \bibinfo {author} {\bibfnamefont {J.}~\bibnamefont {Calarco}},
  \bibinfo {author} {\bibfnamefont {F.}~\bibnamefont {Hersman}}, \bibinfo
  {author} {\bibfnamefont {E.}~\bibnamefont {Jans}}, \bibinfo {author}
  {\bibfnamefont {G.}~\bibnamefont {Kramer}},  \emph {et~al.},\ }\href
  {\doibase 10.1103/PhysRevC.49.955} {\bibfield  {journal} {\bibinfo  {journal}
  {Phys.Rev.}\ }\textbf {\bibinfo {volume} {C49}},\ \bibinfo {pages} {955}
  (\bibinfo {year} {1994})}\BibitemShut {NoStop}%
\bibitem [{\citenamefont {Ajzenberg-Selove}(1991)}]{AjzenbergSelove:1991zz}%
  \BibitemOpen
  \bibfield  {author} {\bibinfo {author} {\bibfnamefont {F.}~\bibnamefont
  {Ajzenberg-Selove}},\ }\href {\doibase 10.1016/0375-9474(91)90446-D}
  {\bibfield  {journal} {\bibinfo  {journal} {Nucl.Phys.}\ }\textbf {\bibinfo
  {volume} {A523}},\ \bibinfo {pages} {1} (\bibinfo {year} {1991})}\BibitemShut
  {NoStop}%
\bibitem [{\citenamefont {Kobayashi}\ \emph {et~al.}(2006)\citenamefont
  {Kobayashi}, \citenamefont {Akimune}, \citenamefont {Ejiri}, \citenamefont
  {Fujimura}, \citenamefont {Fujiwara} \emph {et~al.}}]{Kobayashi:2006gb}%
  \BibitemOpen
  \bibfield  {author} {\bibinfo {author} {\bibfnamefont {K.}~\bibnamefont
  {Kobayashi}}, \bibinfo {author} {\bibfnamefont {H.}~\bibnamefont {Akimune}},
  \bibinfo {author} {\bibfnamefont {H.}~\bibnamefont {Ejiri}}, \bibinfo
  {author} {\bibfnamefont {H.}~\bibnamefont {Fujimura}}, \bibinfo {author}
  {\bibfnamefont {M.}~\bibnamefont {Fujiwara}},  \emph {et~al.},\ }\href@noop
  {} {\enquote {\bibinfo {title} {{De-excitation gamma-rays from the s-hole
  state in N-15 associated with proton decay in O-16}},}\ } (\bibinfo {year}
  {2006}),\ \Eprint {http://arxiv.org/abs/nucl-ex/0604006}
  {arXiv:nucl-ex/0604006 [nucl-ex]} \BibitemShut {NoStop}%
\bibitem [{\citenamefont {Nakahata}\ \emph {et~al.}(1986)\citenamefont
  {Nakahata} \emph {et~al.}}]{Nakahata:1986zp}%
  \BibitemOpen
  \bibfield  {author} {\bibinfo {author} {\bibfnamefont {M.}~\bibnamefont
  {Nakahata}} \emph {et~al.} (\bibinfo {collaboration} {KAMIOKANDE
  Collaboration}),\ }\href {\doibase 10.1143/JPSJ.55.3786} {\bibfield
  {journal} {\bibinfo  {journal} {J.Phys.Soc.Jap.}\ }\textbf {\bibinfo {volume}
  {55}},\ \bibinfo {pages} {3786} (\bibinfo {year} {1986})}\BibitemShut
  {NoStop}%
\bibitem [{\citenamefont {Hayato}(2002)}]{Hayato:2002sd}%
  \BibitemOpen
  \bibfield  {author} {\bibinfo {author} {\bibfnamefont {Y.}~\bibnamefont
  {Hayato}},\ }\href {\doibase 10.1016/S0920-5632(02)01759-0} {\bibfield
  {journal} {\bibinfo  {journal} {Nucl.Phys.Proc.Suppl.}\ }\textbf {\bibinfo
  {volume} {112}},\ \bibinfo {pages} {171} (\bibinfo {year}
  {2002})}\BibitemShut {NoStop}%
\bibitem [{\citenamefont {Benhar}\ \emph {et~al.}(2005)\citenamefont {Benhar},
  \citenamefont {Farina}, \citenamefont {Nakamura}, \citenamefont {Sakuda},\
  and\ \citenamefont {Seki}}]{Benhar:2005dj}%
  \BibitemOpen
  \bibfield  {author} {\bibinfo {author} {\bibfnamefont {O.}~\bibnamefont
  {Benhar}}, \bibinfo {author} {\bibfnamefont {N.}~\bibnamefont {Farina}},
  \bibinfo {author} {\bibfnamefont {H.}~\bibnamefont {Nakamura}}, \bibinfo
  {author} {\bibfnamefont {M.}~\bibnamefont {Sakuda}}, \ and\ \bibinfo {author}
  {\bibfnamefont {R.}~\bibnamefont {Seki}},\ }\href {\doibase
  10.1103/PhysRevD.72.053005} {\bibfield  {journal} {\bibinfo  {journal}
  {Phys.Rev.}\ }\textbf {\bibinfo {volume} {D72}},\ \bibinfo {pages} {053005}
  (\bibinfo {year} {2005})},\ \Eprint {http://arxiv.org/abs/hep-ph/0506116}
  {arXiv:hep-ph/0506116 [hep-ph]} \BibitemShut {NoStop}%
\bibitem [{\citenamefont {Benhar}\ \emph {et~al.}(1994)\citenamefont {Benhar},
  \citenamefont {Fabrocini}, \citenamefont {Fantoni},\ and\ \citenamefont
  {Sick}}]{Benhar:1994hw}%
  \BibitemOpen
  \bibfield  {author} {\bibinfo {author} {\bibfnamefont {O.}~\bibnamefont
  {Benhar}}, \bibinfo {author} {\bibfnamefont {A.}~\bibnamefont {Fabrocini}},
  \bibinfo {author} {\bibfnamefont {S.}~\bibnamefont {Fantoni}}, \ and\
  \bibinfo {author} {\bibfnamefont {I.}~\bibnamefont {Sick}},\ }\href {\doibase
  10.1016/0375-9474(94)90920-2} {\bibfield  {journal} {\bibinfo  {journal}
  {Nucl.Phys.}\ }\textbf {\bibinfo {volume} {A579}},\ \bibinfo {pages} {493}
  (\bibinfo {year} {1994})}\BibitemShut {NoStop}%
\bibitem [{\citenamefont {Bradford}\ \emph {et~al.}(2006)\citenamefont
  {Bradford}, \citenamefont {Bodek}, \citenamefont {Budd},\ and\ \citenamefont
  {Arrington}}]{Bradford:2006yz}%
  \BibitemOpen
  \bibfield  {author} {\bibinfo {author} {\bibfnamefont {R.}~\bibnamefont
  {Bradford}}, \bibinfo {author} {\bibfnamefont {A.}~\bibnamefont {Bodek}},
  \bibinfo {author} {\bibfnamefont {H.~S.}\ \bibnamefont {Budd}}, \ and\
  \bibinfo {author} {\bibfnamefont {J.}~\bibnamefont {Arrington}},\ }\href
  {\doibase 10.1016/j.nuclphysbps.2006.08.028} {\bibfield  {journal} {\bibinfo
  {journal} {Nucl.Phys.Proc.Suppl.}\ }\textbf {\bibinfo {volume} {159}},\
  \bibinfo {pages} {127} (\bibinfo {year} {2006})},\ \Eprint
  {http://arxiv.org/abs/hep-ex/0602017} {arXiv:hep-ex/0602017 [hep-ex]}
  \BibitemShut {NoStop}%
\bibitem [{\citenamefont {Bernheim}\ \emph {et~al.}(1982)\citenamefont
  {Bernheim}, \citenamefont {Bussiere}, \citenamefont {Mougey}, \citenamefont
  {Royer}, \citenamefont {Tarnowski} \emph {et~al.}}]{Bernheim:1981si}%
  \BibitemOpen
  \bibfield  {author} {\bibinfo {author} {\bibfnamefont {M.}~\bibnamefont
  {Bernheim}}, \bibinfo {author} {\bibfnamefont {A.}~\bibnamefont {Bussiere}},
  \bibinfo {author} {\bibfnamefont {J.}~\bibnamefont {Mougey}}, \bibinfo
  {author} {\bibfnamefont {D.}~\bibnamefont {Royer}}, \bibinfo {author}
  {\bibfnamefont {D.}~\bibnamefont {Tarnowski}},  \emph {et~al.},\ }\href
  {\doibase 10.1016/0375-9474(82)90020-3} {\bibfield  {journal} {\bibinfo
  {journal} {Nucl.Phys.}\ }\textbf {\bibinfo {volume} {A375}},\ \bibinfo
  {pages} {381} (\bibinfo {year} {1982})}\BibitemShut {NoStop}%
\bibitem [{\citenamefont {Fissum}\ \emph {et~al.}(2004)\citenamefont {Fissum}
  \emph {et~al.}}]{Fissum:2004we}%
  \BibitemOpen
  \bibfield  {author} {\bibinfo {author} {\bibfnamefont {K.}~\bibnamefont
  {Fissum}} \emph {et~al.} (\bibinfo {collaboration} {Jefferson Lab Hall A
  Collaboration}),\ }\href {\doibase 10.1103/PhysRevC.70.034606} {\bibfield
  {journal} {\bibinfo  {journal} {Phys.Rev.}\ }\textbf {\bibinfo {volume}
  {C70}},\ \bibinfo {pages} {034606} (\bibinfo {year} {2004})},\ \Eprint
  {http://arxiv.org/abs/nucl-ex/0401021} {arXiv:nucl-ex/0401021 [nucl-ex]}
  \BibitemShut {NoStop}%
\bibitem [{\citenamefont {Rohe}\ \emph {et~al.}(2004)\citenamefont {Rohe},
  \citenamefont {Armstrong}, \citenamefont {Asaturyan}, \citenamefont {Baker},
  \citenamefont {Bueltmann} \emph {et~al.}}]{Rohe:2004dz}%
  \BibitemOpen
  \bibfield  {author} {\bibinfo {author} {\bibfnamefont {D.}~\bibnamefont
  {Rohe}}, \bibinfo {author} {\bibfnamefont {C.}~\bibnamefont {Armstrong}},
  \bibinfo {author} {\bibfnamefont {R.}~\bibnamefont {Asaturyan}}, \bibinfo
  {author} {\bibfnamefont {O.}~\bibnamefont {Baker}}, \bibinfo {author}
  {\bibfnamefont {S.}~\bibnamefont {Bueltmann}},  \emph {et~al.},\ }\href
  {\doibase 10.1103/PhysRevLett.93.182501} {\bibfield  {journal} {\bibinfo
  {journal} {Phys.Rev.Lett.}\ }\textbf {\bibinfo {volume} {93}},\ \bibinfo
  {pages} {182501} (\bibinfo {year} {2004})},\ \Eprint
  {http://arxiv.org/abs/nucl-ex/0405028} {arXiv:nucl-ex/0405028 [nucl-ex]}
  \BibitemShut {NoStop}%
\bibitem [{\citenamefont {Rein}\ and\ \citenamefont
  {Sehgal}(1981)}]{Rein:1980wg}%
  \BibitemOpen
  \bibfield  {author} {\bibinfo {author} {\bibfnamefont {D.}~\bibnamefont
  {Rein}}\ and\ \bibinfo {author} {\bibfnamefont {L.~M.}\ \bibnamefont
  {Sehgal}},\ }\href {\doibase 10.1016/0003-4916(81)90242-6} {\bibfield
  {journal} {\bibinfo  {journal} {Annals Phys.}\ }\textbf {\bibinfo {volume}
  {133}},\ \bibinfo {pages} {79} (\bibinfo {year} {1981})}\BibitemShut
  {NoStop}%
\bibitem [{\citenamefont {Oset}\ \emph {et~al.}(1982)\citenamefont {Oset},
  \citenamefont {Toki},\ and\ \citenamefont {Weise}}]{Oset:1982}%
  \BibitemOpen
  \bibfield  {author} {\bibinfo {author} {\bibfnamefont {E.}~\bibnamefont
  {Oset}}, \bibinfo {author} {\bibfnamefont {H.}~\bibnamefont {Toki}}, \ and\
  \bibinfo {author} {\bibfnamefont {W.}~\bibnamefont {Weise}},\ }\href
  {\doibase 10.1016/0370-1573(82)90123-5} {\bibfield  {journal} {\bibinfo
  {journal} {Phys.Rept.}\ }\textbf {\bibinfo {volume} {83}},\ \bibinfo {pages}
  {281} (\bibinfo {year} {1982})}\BibitemShut {NoStop}%
\bibitem [{\citenamefont {Ashery}\ \emph {et~al.}(1981)\citenamefont {Ashery},
  \citenamefont {Navon}, \citenamefont {Azuelos}, \citenamefont {Walter},
  \citenamefont {Pfeiffer} \emph {et~al.}}]{Ashery:1981tq}%
  \BibitemOpen
  \bibfield  {author} {\bibinfo {author} {\bibfnamefont {D.}~\bibnamefont
  {Ashery}}, \bibinfo {author} {\bibfnamefont {I.}~\bibnamefont {Navon}},
  \bibinfo {author} {\bibfnamefont {G.}~\bibnamefont {Azuelos}}, \bibinfo
  {author} {\bibfnamefont {H.}~\bibnamefont {Walter}}, \bibinfo {author}
  {\bibfnamefont {H.}~\bibnamefont {Pfeiffer}},  \emph {et~al.},\ }\href
  {\doibase 10.1103/PhysRevC.23.2173} {\bibfield  {journal} {\bibinfo
  {journal} {Phys.Rev.}\ }\textbf {\bibinfo {volume} {C23}},\ \bibinfo {pages}
  {2173} (\bibinfo {year} {1981})}\BibitemShut {NoStop}%
\bibitem [{\citenamefont {Jones}\ \emph {et~al.}(1993)\citenamefont {Jones},
  \citenamefont {Ransome}, \citenamefont {Cupps}, \citenamefont {Fergerson},
  \citenamefont {Morris} \emph {et~al.}}]{Jones:1993ps}%
  \BibitemOpen
  \bibfield  {author} {\bibinfo {author} {\bibfnamefont {M.}~\bibnamefont
  {Jones}}, \bibinfo {author} {\bibfnamefont {R.}~\bibnamefont {Ransome}},
  \bibinfo {author} {\bibfnamefont {V.}~\bibnamefont {Cupps}}, \bibinfo
  {author} {\bibfnamefont {R.}~\bibnamefont {Fergerson}}, \bibinfo {author}
  {\bibfnamefont {C.}~\bibnamefont {Morris}},  \emph {et~al.},\ }\href
  {\doibase 10.1103/PhysRevC.48.2800} {\bibfield  {journal} {\bibinfo
  {journal} {Phys.Rev.}\ }\textbf {\bibinfo {volume} {C48}},\ \bibinfo {pages}
  {2800} (\bibinfo {year} {1993})}\BibitemShut {NoStop}%
\bibitem [{\citenamefont {Giannelli}\ \emph {et~al.}(2000)\citenamefont
  {Giannelli}, \citenamefont {Ritchie}, \citenamefont {Applegate},
  \citenamefont {Beck}, \citenamefont {Beck} \emph
  {et~al.}}]{Giannelli:2000zy}%
  \BibitemOpen
  \bibfield  {author} {\bibinfo {author} {\bibfnamefont {R.}~\bibnamefont
  {Giannelli}}, \bibinfo {author} {\bibfnamefont {B.}~\bibnamefont {Ritchie}},
  \bibinfo {author} {\bibfnamefont {J.}~\bibnamefont {Applegate}}, \bibinfo
  {author} {\bibfnamefont {E.}~\bibnamefont {Beck}}, \bibinfo {author}
  {\bibfnamefont {J.}~\bibnamefont {Beck}},  \emph {et~al.},\ }\href {\doibase
  10.1103/PhysRevC.61.054615} {\bibfield  {journal} {\bibinfo  {journal}
  {Phys.Rev.}\ }\textbf {\bibinfo {volume} {C61}},\ \bibinfo {pages} {054615}
  (\bibinfo {year} {2000})}\BibitemShut {NoStop}%
\bibitem [{\citenamefont {Engelhardt}\ \emph {et~al.}(1976)\citenamefont
  {Engelhardt}, \citenamefont {Lewis},\ and\ \citenamefont
  {Ullrich}}]{Engelhardt:1975ct}%
  \BibitemOpen
  \bibfield  {author} {\bibinfo {author} {\bibfnamefont {H.}~\bibnamefont
  {Engelhardt}}, \bibinfo {author} {\bibfnamefont {C.}~\bibnamefont {Lewis}}, \
  and\ \bibinfo {author} {\bibfnamefont {H.}~\bibnamefont {Ullrich}},\ }\href
  {\doibase 10.1016/0375-9474(76)90486-3} {\bibfield  {journal} {\bibinfo
  {journal} {Nucl.Phys.}\ }\textbf {\bibinfo {volume} {A258}},\ \bibinfo
  {pages} {480} (\bibinfo {year} {1976})}\BibitemShut {NoStop}%
\bibitem [{\citenamefont {Abe}\ \emph {et~al.}(2013{\natexlab{c}})\citenamefont
  {Abe} \emph {et~al.}}]{Abe:2013fuq}%
  \BibitemOpen
  \bibfield  {author} {\bibinfo {author} {\bibfnamefont {K.}~\bibnamefont
  {Abe}} \emph {et~al.} (\bibinfo {collaboration} {T2K Collaboration}),\ }\href
  {\doibase 10.1103/PhysRevLett.111.211803} {\bibfield  {journal} {\bibinfo
  {journal} {Phys.Rev.Lett.}\ }\textbf {\bibinfo {volume} {111}},\ \bibinfo
  {pages} {211803} (\bibinfo {year} {2013}{\natexlab{c}})},\ \Eprint
  {http://arxiv.org/abs/1308.0465} {arXiv:1308.0465 [hep-ex]} \BibitemShut
  {NoStop}%
\bibitem [{\citenamefont {Abe}\ \emph {et~al.}(2011{\natexlab{c}})\citenamefont
  {Abe} \emph {et~al.}}]{Abe:2010hy}%
  \BibitemOpen
  \bibfield  {author} {\bibinfo {author} {\bibfnamefont {K.}~\bibnamefont
  {Abe}} \emph {et~al.} (\bibinfo {collaboration} {Super-Kamiokande
  Collaboration}),\ }\href {\doibase 10.1103/PhysRevD.83.052010} {\bibfield
  {journal} {\bibinfo  {journal} {Phys.Rev.}\ }\textbf {\bibinfo {volume}
  {D83}},\ \bibinfo {pages} {052010} (\bibinfo {year} {2011}{\natexlab{c}})},\
  \Eprint {http://arxiv.org/abs/1010.0118} {arXiv:1010.0118 [hep-ex]}
  \BibitemShut {NoStop}%
\bibitem [{\citenamefont {Smy}(2008)}]{Smy:icrc}%
  \BibitemOpen
  \bibfield  {author} {\bibinfo {author} {\bibfnamefont {M.}~\bibnamefont
  {Smy}},\ }in\ \href@noop {} {\emph {\bibinfo {booktitle} {{Proceedings of the
  30th International Cosmic Ray Conference}}}},\ Vol.~\bibinfo {volume} {5},\
  \bibinfo {editor} {edited by\ \bibinfo {editor} {\bibfnamefont
  {R.}~\bibnamefont {Caballero}}, \bibinfo {editor} {\bibfnamefont {J.~C.}\
  \bibnamefont {D'Olivo}}, \bibinfo {editor} {\bibfnamefont {G.}~\bibnamefont
  {Medina-Tanco}}, \bibinfo {editor} {\bibfnamefont {L.}~\bibnamefont
  {Nellen}}, \bibinfo {editor} {\bibfnamefont {F.~A.}\ \bibnamefont
  {Sánchez}}, \ and\ \bibinfo {editor} {\bibfnamefont {J.~F.}\ \bibnamefont
  {Vald\'{e}s-Galicia}}}\ (\bibinfo {address} {Universidad Nacional
  Aut\'{o}noma de M\'{e}xico, Mexico City},\ \bibinfo {year} {2008})\ pp.\
  \bibinfo {pages} {1279--1282}\BibitemShut {NoStop}%
\bibitem [{\citenamefont {Abe}\ \emph {et~al.}(2014)\citenamefont {Abe},
  \citenamefont {Hayato}, \citenamefont {Iida}, \citenamefont {Iyogi},
  \citenamefont {Kameda} \emph {et~al.}}]{Abe:2013gga}%
  \BibitemOpen
  \bibfield  {author} {\bibinfo {author} {\bibfnamefont {K.}~\bibnamefont
  {Abe}}, \bibinfo {author} {\bibfnamefont {Y.}~\bibnamefont {Hayato}},
  \bibinfo {author} {\bibfnamefont {T.}~\bibnamefont {Iida}}, \bibinfo {author}
  {\bibfnamefont {K.}~\bibnamefont {Iyogi}}, \bibinfo {author} {\bibfnamefont
  {J.}~\bibnamefont {Kameda}},  \emph {et~al.},\ }\href {\doibase
  10.1016/j.nima.2013.11.081} {\bibfield  {journal} {\bibinfo  {journal}
  {Nucl.Instrum.Meth.}\ }\textbf {\bibinfo {volume} {A737}},\ \bibinfo {pages}
  {253} (\bibinfo {year} {2014})},\ \Eprint {http://arxiv.org/abs/1307.0162}
  {arXiv:1307.0162 [physics.ins-det]} \BibitemShut {NoStop}%
\bibitem [{\citenamefont {Cravens}\ \emph {et~al.}(2008)\citenamefont {Cravens}
  \emph {et~al.}}]{Cravens:2008aa}%
  \BibitemOpen
  \bibfield  {author} {\bibinfo {author} {\bibfnamefont {J.}~\bibnamefont
  {Cravens}} \emph {et~al.} (\bibinfo {collaboration} {Super-Kamiokande
  Collaboration}),\ }\href {\doibase 10.1103/PhysRevD.78.032002} {\bibfield
  {journal} {\bibinfo  {journal} {Phys.Rev.}\ }\textbf {\bibinfo {volume}
  {D78}},\ \bibinfo {pages} {032002} (\bibinfo {year} {2008})},\ \Eprint
  {http://arxiv.org/abs/0803.4312} {arXiv:0803.4312 [hep-ex]} \BibitemShut
  {NoStop}%
\bibitem [{\citenamefont {Abe}\ \emph {et~al.}(2011{\natexlab{d}})\citenamefont
  {Abe} \emph {et~al.}}]{Abe:2011sj}%
  \BibitemOpen
  \bibfield  {author} {\bibinfo {author} {\bibfnamefont {K.}~\bibnamefont
  {Abe}} \emph {et~al.} (\bibinfo {collaboration} {T2K Collaboration}),\ }\href
  {\doibase 10.1103/PhysRevLett.107.041801} {\bibfield  {journal} {\bibinfo
  {journal} {Phys.Rev.Lett.}\ }\textbf {\bibinfo {volume} {107}},\ \bibinfo
  {pages} {041801} (\bibinfo {year} {2011}{\natexlab{d}})},\ \Eprint
  {http://arxiv.org/abs/1106.2822} {arXiv:1106.2822 [hep-ex]} \BibitemShut
  {NoStop}%
\bibitem [{\citenamefont {Ejiri}(1993)}]{Ejiri:1993rh}%
  \BibitemOpen
  \bibfield  {author} {\bibinfo {author} {\bibfnamefont {H.}~\bibnamefont
  {Ejiri}},\ }\href {\doibase 10.1103/PhysRevC.48.1442} {\bibfield  {journal}
  {\bibinfo  {journal} {Phys.Rev.}\ }\textbf {\bibinfo {volume} {C48}},\
  \bibinfo {pages} {1442} (\bibinfo {year} {1993})}\BibitemShut {NoStop}%
\bibitem [{\citenamefont {Nakahata}\ \emph {et~al.}(1999)\citenamefont
  {Nakahata} \emph {et~al.}}]{Nakahata:1998pz}%
  \BibitemOpen
  \bibfield  {author} {\bibinfo {author} {\bibfnamefont {M.}~\bibnamefont
  {Nakahata}} \emph {et~al.} (\bibinfo {collaboration} {Super-Kamiokande
  Collaboration}),\ }\href {\doibase 10.1016/S0168-9002(98)01200-5} {\bibfield
  {journal} {\bibinfo  {journal} {Nucl.Instrum.Meth.}\ }\textbf {\bibinfo
  {volume} {A421}},\ \bibinfo {pages} {113} (\bibinfo {year} {1999})},\ \Eprint
  {http://arxiv.org/abs/hep-ex/9807027} {arXiv:hep-ex/9807027 [hep-ex]}
  \BibitemShut {NoStop}%
\bibitem [{\citenamefont {Adamson}\ \emph {et~al.}(2011)\citenamefont {Adamson}
  \emph {et~al.}}]{Adamson:2011ku}%
  \BibitemOpen
  \bibfield  {author} {\bibinfo {author} {\bibfnamefont {P.}~\bibnamefont
  {Adamson}} \emph {et~al.} (\bibinfo {collaboration} {MINOS Collaboration}),\
  }\href {\doibase 10.1103/PhysRevLett.107.011802} {\bibfield  {journal}
  {\bibinfo  {journal} {Phys.Rev.Lett.}\ }\textbf {\bibinfo {volume} {107}},\
  \bibinfo {pages} {011802} (\bibinfo {year} {2011})},\ \Eprint
  {http://arxiv.org/abs/1104.3922} {arXiv:1104.3922 [hep-ex]} \BibitemShut
  {NoStop}%
\end{thebibliography}%

\end{document}